\documentclass{emulateapj}
\usepackage{epstopdf}
\usepackage{subfigure, amsmath, mathtools}

\newcommand{\feh}{\hbox{$ [{\rm Fe}/{\rm H}]$ }} 
\def\Sec{${}^{\prime\prime}$\llap{.}}

\slugcomment{AJ accepted}

\shorttitle{NGC$\,$6656 Horizontal Branch}
\shortauthors{Kunder et al.}

\begin{document}

\title{The RR Lyrae variables and Horizontal Branch of NGC$\,$6656 (M22)}\thanks{Based in part on observations made with the European Southern Observatory (ESO) telescopes and obtained from the ESO/ST-ECF Science Archive facility.}\thanks{This research draws upon data distributed by the NOAO Science Archive. NOAO is operated by the Association of Universities for Research in Astronomy (AURA) under cooperative agreement with the National Science Foundation.}

\author{Andrea Kunder\altaffilmark{1},
Peter B. Stetson\altaffilmark{2}, 
Santi Cassisi\altaffilmark{3},
Andrew Layden\altaffilmark{4},
Giuseppe Bono\altaffilmark{5,6},
M\'{a}rcio Catelan\altaffilmark{7,8},
Alistair R. Walker\altaffilmark{1},
Leonardo Paredes Alvarez\altaffilmark{1},
James L. Clem\altaffilmark{9},
Noriyuki Matsunaga\altaffilmark{10},
Maurizio Salaris\altaffilmark{11},
Jae-Woo Lee\altaffilmark{12},
Brian Chaboyer\altaffilmark{13}
}

\altaffiltext{1}{Cerro Tololo Inter-American Observatory, Casilla 603, La Serena, Chile}
\affil{E-mail: akunder@ctio.noao.edu}
\altaffiltext{2}{ Dominion Astrophysical Observatory, NRC-Herzberg, National Research Council, Victoria BC, Canada V9E~2E7}
\altaffiltext{3}{INAF-Osservatorio Astronomico di Collurania, Via M. Maggini, I-64100 Teramo, Italy}
\altaffiltext{4}{Bowling Green State University, Bowling Green, OH 43403, USA }
\altaffiltext{5}{Dipartimento di Fisica, Universita di Roma Tor Vergata, Rome, Italy}
\altaffiltext{6}{INAF--Osservatorio Astronomico di Roma, via Frascati 33 00040 Monte Porzio Catone, Italy}
\altaffiltext{7}{Pontificia Universidad Cat\'olica de Chile, Departamento de Astronom\'\i a y Astrof\'\i sica, Av. Vicu\~{n}a Mackenna 4860, 782-0436 Macul, Santiago, Chile; e-mail: mcatelan@astro.puc.cl}  
\altaffiltext{8}{The Milky Way Millennium Nucleus, Av. Vicu\~{n}a Mackenna 4860, 782-0436 Macul, Santiago, Chile} 
\altaffiltext{9}{Department of Physics and Astronomy, Louisiana State University, Baton Rouge, LA 70803-4001}
\altaffiltext{10}{Department of Astronomy, School of Science, The University of Tokyo, Japan}
\altaffiltext{11}{Astrophysics Research Institute, Liverpool John Moores University, Twelve Quays House, Egerton Wharf, Birkenhead CH41 1LD, UK}
\altaffiltext{12}{Korea Astronomy and Space Science Institute, Daejeon 305-348, Republic of Korea}
\altaffiltext{13}{Department of Physics and Astronomy, Dartmouth College, Hanover, NH 03755, USA}

\begin{abstract}
The first calibrated broadband {\it UBVI\/} time-series photometry is presented for the RR Lyrae
variable stars in NGC$\,$6656 (M22), with observations spanning a range of twenty-two years.  
We have also redetermined the variability types and periods for the RR Lyrae stars
identified previously by photographic observations, revising the number of fundamental-mode 
RR Lyrae variables (RR0) to 10 and the number of first-overtone variables (RR$1$) 
to 16.  The mean periods of the RR0 and RR$1$ variables are 
$\langle P \rangle_{RR0}=$0.66$\pm$0.02 d and $\langle P \rangle_{RR1}=$0.33$\pm$0.01 d, respectively, 
supporting an Oosterhoff II classification for the cluster.  The number ratio 
of RR$1$- to all RR-type variables is $N_1/N_{RR}$=0.61, also consistent with an Oosterhoff II designation.  
Both the RR Lyrae stars' minimum light colors and the blue
edge of the RR Lyrae instability strip suggest $E(\hbox{\it B--V\/})$=0.36$\pm$0.02 mag toward M22.
Regarding the HB morphology of M22, we find 
($B$-$R$)/($B$+$V$+$R$)=+0.97$\pm$0.1 and at least one ``gap" located in an unusual part
of the blue HB, in the middle of the so-called hot HB stars.  
%
\end{abstract}


\section{Introduction}

NGC 6556 (M22) is a relatively massive galactic globular cluster (GC) with $M_V = -8.50$, lying inside the solar
circle with ($l$,$b$)=(9.89,$-$7.55) and in relatively close proximity to the Sun, $\rm R_{\sun}$ = 3.2 kpc.
These and other basic parameters are listed in \citep[][2010 edition]{harris96}.  Of note is that 
as far as massive GCs are concerned, M22 is not as heavily crowded
as some, with a core radius of $r_c$=1.33 arcmin, a central luminosity
density of $\rho_0$=3.63 $\rm L_{\sun}~pc^{-3}$ \citep[2010 edition of][]{harris96}
and a tidal radius of $r_t$= 27 pc \citep{mackey05}. Thus despite quite high reddening in which there is likely
to be some variation, $E(\hbox{\it B--V\/})$ = 0.34 mag, M22 is amenable to ground-based observations.

M22 was first recognized as peculiar when 
it was discovered that its color-magnitude diagram (CMD) is 
anomalous:  it has a red giant branch (RGB) with a shallower slope than would otherwise 
be expected based on other metallicity indicators \citep{butler73, hesser76, hesser77, hesserhart79}.  
Also, importantly, \citet{hesser77} found that M22 had a large color spread in the RGB, and 
therefore linked M22 to the prototypical peculiar GC, $\omega$ Cen.  
Since then, detailed spectroscopic investigations have shown that, like $\omega$ Cen, M22 
shows variations in its
bulk heavy-element content, including iron and elements associated with slow neutron-capture processes
($s$-elements) \citep[e.g.,][]{norris83, dacosta09, marino09, marino12}.

In particular, from high resolution spectra ($R$$\sim$38 000 to $R$$\sim$60 000) of 35 
bright giant stars, \citet{marino11} have shown that the stars in M22 exhibit 
a large range in their abundance of $s$-elements,in particular the $\rm [La/Eu]$ ratio 
cleanly divides the sample into two different sub-populations: an 
$s$-rich stellar component and an $s$-poor one.  This was then later 
also seen in 101 SGB stars using medium-resolution ($R$$\sim$6400) 
spectra \citep{marino12}.  There is additionally a strong correlation between $\rm [Fe/H]$ and
$s$-element abundance; the $s$-rich stars have a systematically higher $\rm [Fe/H]$
with ${\rm \langle[Fe/H]\rangle_{s-rich}=-1.68\pm0.02}$, whereas the $s$-poor stars are 
found to have ${\rm \langle[Fe/H]\rangle_{s-poor}=-1.82\pm0.02}$ \citep{marino09}.  
This is a ${\rm \Delta[Fe/H]}$ of 0.14$\pm$0.03 dex.  The CNO-sum distribution was 
also found to differ between the $s$-rich and $s$-poor stellar components, on the 
order of ${\rm \Delta[C+N+O/Fe]}$$\sim$0.13~dex, with the $s$-rich one being the 
CNO-enhanced component \citep{marino09, marino11}.  Therefore, M22 is one of 
the few GCs showing clear evidence of a bimodality in the CNO-sum distribution 
\citep[as is also seen in $\omega$~Cen, e.g.,][]{marino12omega}.

GCs are generally chemically homogeneous when it comes to the abundances of the 
iron-peak elements \citep{kraft03}.  The well-established counterexamples, such as 
$\omega$ Centauri \citep{freeman75, norris95, pancino02},  
M54 in the Sagittarius dSph \citep{bellazzini08, carretta10a, saviane12},  
NGC$\,$2419 \citep{cohen10, mucciarelli12},  
NGC$\,$1851 \citep{carretta10b},  
M22 \citep{dacosta09, marino09} and  
Terzan$\,$5 \citep{ferraro09}
may have a different origin than the other GCs.  For example, they could have an extragalactic 
origin, such as being the remnant of a dwarf galaxy tidally disrupted by the MW.  That these GCs have a 
peculiar origin is also suggested because they are some of the most massive GCs in the Galaxy and have
different kinematics (e.g., retrograde rotation) than many ``normal" GCs \citep{lee07}.

Photometrically, the two distinct sub-populations present in M22 have been traced 
along the RGB using Str\"{o}mgren filters \citep{milone12}, and along the SGB using optical 
 {\it Hubble Space Telescope (HST)} filters \citep{piotto12}.  The interpretative analysis 
 provided by \citet{cassisi08} and \citet{sbordone11} suggests that the observed split along 
 these evolutionary sequences can be understood as due to the bimodality in the CNO-sum 
 and the observed light-element anti-correlations, respectively, the latter now understood to be a 
 ubiquitous feature of GCs  \citep{carretta09a, carretta09b}

The spectroscopic study of multiple populations in Galactic GCs has recently been extended 
to the HB,  
e.g., M4, 
\citet{marino11m4}; NGC$\,$1851, \citet{gratton12}; NGC$\,$2808, \citet{gratton11}
and  47~Tuc, \citet{milone1247tuc}, \citet{gratton13}.  
On the basis of these measurements a scenario is emerging in which 
Na-poor/O-rich stars, the first generation progeny, are located in the red portion of the HB, whereas 
Na-rich/O-poor stars, the stars belonging to the second generation, are distributed along the 
blue portion of the HB.  Since the presence of a Na-O anticorrelation has to be accompanied 
by (at least a moderate) helium enhancement, and He-rich stars are expected on theoretical 
grounds to populate the bluest portion of the HB distribution, these empirical findings provide 
plain evidence that the HB morphology is greatly affected by the multiple-population phenomenon.
In the case of NGC$\,$1851, spectroscopy supplemented with an accurate analysis of both the HB 
morphology and the pulsation properties of RR Lyrae stars \citep{kunder13a} has been used to 
shed more light on the detailed distribution of the various sub-populations along the HB of this 
peculiar GC.  Very recently, the first spectroscopic tracing of sub-populations along the 
HB of M22 has been carried out by Marino, Milone \& Lind (2013), finding that all the (seven) 
red HB stars (including 3 RR Lyrae variables) analyzed are barium-poor and sodium-poor and 
belong to the first generation of stars.

In this paper we present new {\it UBVI\/} photometry of M22, based on original and archival observations,
which has allowed a more detailed study of the RR Lyrae instability strip than has previously 
been possible, including the discovery of 
additional RR Lyrae stars and the first calibrated light 
curves of the complete M22 RR Lyrae variable sample.  We will use these data 
to seek evidence of the metallicity and generational dichotomies described above 
among the RRL in M22.
\section{Observations}
The observations come from co-author Peter B. Stetson's (PBS) data archive and the details
are given in Table~1.  The last column in the table with the heading ``Multiplex" refers
to the number of individual CCDs in the camera used for the observing run.
Standard DAOPHOT/ALLFRAME procedures \citep{ste87, ste90, ste94} 
were used to perform profile-fitting photometry with aperture growth-curve corrections.  
The calibration of the instrumental data is to the Johnson/Kron-Cousins $UBVI$
photometric system defined by \citep{landolt92}, and was carried out as described by \citet{ste00, stetson05}.
The observations are contained within 61 datasets, where a dataset is either the complete 
body of data from one CCD on one photometric night or the body of data 
from one CCD on one or more non-photometric nights during the same observing run.  
Therefore the number of datasets and the number of images will be different, as datasets,
individually calibrated to the Landolt system, may be broken up to compensate for e.g., changing
weather conditions.  Of the 61 datasets, 35 are considered photometric, and the remaining 26 
are considered non-photometric.  The 35 photometric datasets are used to set up local standards
in the field of M22, using nightly calibration equations that include linear and quadratic color terms as well 
as linear extinction terms; a color-extinction term is also employed
for the $B$ filter.  These local standards are then used to determine the photometric
zero points of individual CCD images obtained on non-photometric nights.

The maximum number of calibrated magnitude measurements for any one star is 24 in $U$, 154 
in $B$, 206 in $V$, and 44 in $I$.  There was one photometric night in which $R$-band 
data was obtained, but we consider that insufficient for a reliable photometric calibration, and a discussion of
M22 $R$-band photometry is not presented here.  However, the $R$-band images were included 
in the ALLFRAME reductions to aid in the completeness and astrometric precision of the catalog.
Our final catalog includes 620,730 objects with photometric measurements and spans a field 
roughly 38.5 arcminutes east-west by 49.2 arcminutes north-south.  There are 525,585 objects 
with ``useful'' measurements (arbitrarily defined as $\sigma$(magnitude) $<$ 0.10 mag) in 
$B$, $V$, and $I$; 55,910 of these also have useful (same definition) photometry in $U$.  
There are a further 96,857 objects with astrometric information only (46.8 by 49.2 arcminutes).  

The astrometry is carried out as in \citet{kunder13b}, tied to the U.S. Naval Observatory (USNO) 
A2.0 Astrometric Reference Catalog.  Therefore we believe our positions are on the USNO-A2.0 system
with an accuracy well 
within 0\Sec1, and are internally precise to better than 0\Sec03.
%
%
\clearpage
\begin{table}
\begin{scriptsize}
\centering
\caption{M22 Observations}
\label{obs}
\begin{tabular}{lllcccclc} \hline
Run ID & Dates & Telescope/Camera/Detector  & $U$ & $B$ & $V$ & $R$ & $I$ & Multiplex \\ \hline
\hline
 1  bond21      &    1991 Sep 23       &       CTIO 0.9m          772         &       --       &   2       &   2      &   --      &   --     & \\
 2  emmi5        &   1993 Jul 17--18      &    ESO NTT 3.6m    EMMI        &       --        & 20      &   21      &   --      &   --     &  \\
 3  apr97         &   1997 Apr 16        &      ESO Dutch 0.9m  Tektronix $33$       &      --      &   --       &   8      &   --       &   8     &  \\
 4  bond6        &    1998 Apr 22       &       CTIO 0.8m       Tek2K\_3       &      1       &   1       &   1       &  --        &  1     &  \\
 5  dmd           &   1998 Jun 25       &       JKT 1.0m        TEK4          &     --      &   --       &   2      &   --      &    3     & \\
 6  wfi9            &  1999 May 15        &      ESO/MPI 2.2m    WFI        &        --     &     4       &   3     &    --       &   3  &   $\times$8 \\
 7  wfi12           &   1999 Jul 12       &       ESO/MPI 2.2m    WFI        &        --      &    1      &    4     &    --      &    2   &  $\times$8 \\
 8  wfi10           &   2000 Jul 07       &       ESO/MPI 2.2m    WFI          &      --      &    4      &    4      &   --      &    3   &  $\times$8 \\
 9  ct36aug00  & 2000 Aug 30--31      &    CTIO 0.9m       Tek2K\_3       &     --      &   78      &   78       &  --       &  -- &  \\
10  wfi5            &   2002 Jun 18        &      ESO/MPI 2.2m    WFI           &     --      &    6       &   6       &  --      &    6   &  $\times$8 \\
11  susi03may      &  2003 May 31     &         ESO NTT 3.6m    SUSI       &         8       &  --        &  8      &   --      &   --    & $\times$2 \\
12  wfi26           &   2004 Jun 26      &        ESO/MPI 2.2m    WFI          &      --        &  4      &    6       &  --      &   --       &    $\times$8 \\
13  fors20605   &   2006 May 29     &         ESO VLT 8.0m    FORS2         &     --       &  --        &  3        &  5       &   3      &     $\times$2 \\
14  fors0707     &    2007 Jul 03--05     &     ESO VLT 8.0m    FORS1         &      1      &   --      &   38      &   --       &  --      &     $\times$2 \\
15  efosc09      &     2009 Apr 20--29     &     ESO NTT 3.6m    EFOSC LORAL       & 16      &   69      &   10       &   3       &  --     &  \\
16  ct12aug     &     2012 Aug 19--21     &     CTIO 0.9m       Tek2K\_3        &    20       &  15      &   45      &   --       &  15     &  \\
\hline
\hline
\end{tabular}
Notes: \\
 1 Observer H.~E.~Bond
 2 Observers ``SAV/ZAGGIA''
 3 Observer A.~Rosenberg?
 4 Observer H.~E.~Bond
 5 Observer ``DMD''
 6 Program identification 163.O-0741(C)
 7 Program identification unknown, observer unknown
 8 Program identification 065.L-0463, observer Ferraro
 9 Observers A.~Walker \& D.~Walker
10 Program identification 69.D-0582(A)
11 Program identification 71.D-0175(A)
12 Program identification 073.D-0188(A)
13 Program identification 077.D-0775(A)
14 Program identification 079.D-0893(A)
15 Program identification 083.D-0544(A)
16 Proposal ID 2012B-0178, observers A.~Kunder, L. Paredes Alvarez
\end{scriptsize}
  \end{table}

\clearpage
\begin{table}
\begin{scriptsize}
\centering
\caption{RR Lyrae stars in M22}
\label{lcpars}
\begin{tabular}{p{0.35in}p{0.6in}p{0.6in}p{0.5in}p{0.25in}p{0.25in}p{0.25in}p{0.25in}p{0.2in}p{0.2in}p{0.2in}p{0.2in}p{0.3in}p{0.25in}p{0.4in}p{0.5in}} \\ \hline
Name & R.A. (J2000.0) & Decl. (J2000.0) & Period (d) & $\langle$$U$$\rangle$ & $\langle$$B$$\rangle$ & $\langle$$V$$\rangle$ & $\langle$$I$$\rangle$ & $A_U$ & $A_B$ & $A_V$ & $A_I$ & Type & $R_{proj}$ (') & Separation Index & Comment \\ 
\hline
V1 & 18 36 19.55 & $-$23 54 32.7 & 0.615541  &  15.15 &  15.00 &  14.27 &  13.26 &  1.35 &  1.48 &  1.15 &  0.69 & RR0 &  1.1 & 5.9 & period increasing?\\ 
V2 & 18 36 34.90 & $-$23 53 06.3 & 0.641718 & 14.81  & 14.79 &  14.10 &  13.19 &  1.16 &  1.21 &  0.93 &  0.57 & RR0 & 3.0 & 5.5 & \\ %
V3 & 18 36 37.99 & $-$23 47 14.9 & 0.539559: & -- & 16.49 & 15.64 & 14.69 & -- & -- & -- & --  & RR0 & 7.1 & & $^a$  \\
V4 & 18 36 23.35 & $-$23 55 29.1 & 0.716393  & 15.09 &  14.94 &  14.17 &  13.15 &  1.02 &  0.98 &  0.80 &  0.52 & RR0 &  1.2 & 4.8 & \\
V6 & 18 36 18.24 & $-$23 56 03.4 & 0.638486 &  14.91  & 14.80  & 14.10  & 13.14  & 1.35  & 1.37  & 1.10 &  0.70 & RR0 &  2.3 & 6.3 & $^b$  \\
V7 & 18 35 57.60 & $-$23 47 40.5 &  0.649523 &  14.94 &  14.81 &  14.10 &  13.20 &  1.19 &  1.44 &  1.26 &  0.70 & RR0 & 9.3 &  10.7 & \\
V10 & 18 36 20.92 & $-$23 56 27.1 & 0.646028  & 14.93  & 14.82  & 14.13  & 13.16  & 1.05  & 1.41 &  1.15 &  0.64 & RR0 & 2.3 & 6.1 &  \\ %
V12 & 18 36 23.75  & $-$23 55 38.8 &  0.322622  & 15.00  & 14.78  & 14.19  & 13.38  & 0.61 & 0.57 & 0.44 & 0.27 & RR$1$ & 1.4 & 5.6 &  \\ 
V13 & 18 36 28.64   & $-$23 51 39.5  & 0.672530  & 14.87  & 14.74 &  14.06 &  13.14 &  1.10 &  1.36 &  1.08 &  0.65 & RR0 & 2.9 & 6.5 & \\ %
V15 & 18 36 32.07   & $-$23 55 40.6  & 0.370922 & 14.98  & 14.84  & 14.21  & 13.34  & 0.49  & 0.56  & 0.41  & 0.23  & RR$1$  & 2.5 & 6.2 &  \\ 
V16 & 18 36 36.97   & $-$23 54 33.2  & 0.325293  & 15.06  & 14.80  & 14.24  & 13.44  & 0.49  & 0.54  & 0.44  & 0.24   & RR$1$ & 3.3 & 5.6 &  \\ 
V18 & 18 36 16.14   & $-$23 47 14.9  & 0.321059 & -- &  14.65 & 14.11 & 13.35 & -- & 0.57  & 0.46  & 0.30:  & RR$1$ & 7.3 & 10.3 & \\ %
V19 & 18 36 20.58   & $-$23 52 17.2 & 0.383621 & 15.06  & 14.75  & 14.19  & 13.33   &0.55  & 0.58  & 0.46  & 0.35   & RR$1$ & 2.2  & 7.0 &  \\ %
V20 & 18 36 14.79   & $-$23 56 32.9  & 0.756134  & 14.98  & 14.84 &  14.10 &  13.09 &  0.80 &  0.93 &  0.70 &  0.43 & RR0 & 3.2 &  6.1 & \\ %
V21 & 18 36 25.74  & $-$23 52 57.9 & 0.327134  & 14.73  & 14.64  & 14.07  & 13.30  & 0.42  & 0.54  & 0.40  & 0.23  & RR$1$ & 1.4  & 5.2 &  \\ %
V22* & 18 35 03.8 & $-$23 50 58.2 & 0.6245374 & -- & -- & 13.17 & -- & -- & -- & 0.41 & -- & field RR0 & 20.3 & & V3853 Sgr \\
V23  & 18 36 23.01   & $-$23 54 41.4  & 0.581019:  & 15.36  & 14.93  & 14.28 &  13.32 &  1.25 &  1.37 &  1.02 &  0.62 & RR0 & 0.5 & 5.2 & \\ %
V24 &  18 36 24.04  &  $-$23 54 29.4 & -- & 14.1 & 13.1 & 11.3 & 9.3 & -- & -- & -- & -- & NV & 0.4 & & not a variable \\
V25  & 18 36 46.29   & $-$23 48 02.2 & 0.399985  & 15.02  & 14.83  & 14.18  & 13.30  & 0.47  & 0.62  & 0.46  & 0.31  & RR$1$ & 8.4 & 9.5 & Blazhko? \\
V27* & 18 35 28.92  & $-$23 45 25.3 & 0.34278 & -- & -- & 13.33 & -- & -- & -- & 0.32 & -- & field RR$1$ & 16.4 & & V2592 Sgr \\
V29 & 18 36 30.85 & $-$23 41 03.2 & 0.471584 & -- & 14.79 & 14.21 & 13.37 & -- & 0.45 & 0.36 & -- & field RR$1$ & 13.3 & & NSV 11080 \\
KT-12  & 18 36 30.93  & $-$23 53 48.8 & 0.443610  & --  & 17.28  & 16.56  & 15.59  & --  & 1.18  & 0.93  & 0.50 & RR0 & 1.8 & & bulge star \\
KT-14  & 18 36 30.67  & $-$23 53 53.8 & 0.371984  & 14.85  & 14.69  & 14.07  & 13.25  & 0.39  & 0.35  & 0.30  & 0.18  & RR$1$ & 1.7 &  4.3 & Blazhko? blending? \\
KT-16 & 18 36 30.36  & $-$23 57 12.9 & 0.2819       & 14.95  &  14.71  &  14.17  &  13.45  &  0.07  &  0.11 &  0.07  & 0.05 & RR$1$ & 3.3 & 6.6 & \\ 
KT-26 & 18 36 23.15  & $-$23 53 23.3 &  0.370960 &  15.00 & 14.74 & 14.12 & 13.29 & 0.17 & 0.32 & 0.17 & 0.14 & RR$1$ & 0.9 & 2.1 &  \\
KT-36  & 18 36 15.88  & $-$23 56 06.9 & 0.315182 & 14.99  & 14.75  & 14.17  & 13.41  & 0.36  & 0.46  & 0.37  & 0.22  & RR$1$ & 2.7 & 7.2 & \\
KT-37  & 18 36 13.17  & $-$23 53 46.8 & 0.296058  & 14.94   & 14.72  & 14.16  & 13.42  & 0.14  & 0.17  & 0.13  & 0.07  & RR$1$  & 2.7 & 6.2 & \\ %
KT-55  & 18 36 23.23  & $-$23 53 57.9 & 0.658735  & 15.07  & 14.85  & 14.14  & 13.17  & 0.86  & 1.3  & 0.75  & 0.4: & RR0 & 0.4 & 4.1 & Blazhko? \\
\hline
NV1 &  18 35 59.11 & $-$23 57 13.1 & 0.305811  & 14.92  & 14.66  & 14.14  & 13.41  & 0.24  & 0.28  & 0.22  & 0.11   & RR$1$ & 6.9 & 8.7 & \\ 
NV2  &  18 36 02.96 & $-$23 50 29.4 & 0.332917:  & 99.99  & 14.83  & 14.23  & 13.45  & 9.99  & 0.25  & 0.16  & 0.15: & RR$1$ & 6.5 & 3.1 & period uncertain \\ 
NV3 & 18 36 29.52 & $-$24 01 32.6 & 0.334020  & 99.99  & 14.58  & 14.05  & 13.38  & 9.99  & 0.48  & 0.40  &  --     & RR$1$ & 7.4 & 5.4 & \\ 
NV4  &  18 36 31.67 & $-$23 49 30.3 & 0.287082  & 15.08  & 14.72  & 14.18  & 13.42  & 0.35  & 0.16  & 0.11  & 0.08 & RR$1$ & 5.2 & 8.5 & \\ 
\hline
\end{tabular}
\end{scriptsize}
* insufficient observations (star is outside field of view), data presented here are taken from the ASAS survey \citet{pojmanski00}\\
$^a$ inadequate phase coverage, too faint to be cluster RRL \\
$^b$ some light curve instability seen, e.g., phase jumps
\end{table}
\clearpage


%
%
\section{Results}
\subsection{RR Lyrae variables}
The first variable stars (V1-16) in M22 were discovered more than a century 
ago by \citet{bailey02}.  Gradually additional variables were discovered: V17 
in \citet{shapley27}, V18-25 by \citet{sawyer44} and V26-31 by \citet{hoffleit72}.  
\citet{wehlau77} published the discovery of V32 and V33, \citet{lloydevans78} 
published the discovery of V34 and V35, and \citet{kravtsov94} identified
V36-43, although the membership of these variables is uncertain.
Finally the first results from CCD observations were presented by \citet{kaluzny01}
resulting in 36 new variables.  They were not able to determine periods for the seven 
RR Lyrae variables they detected, and provide approximate $B$ and $V$ calibration 
of their photometry.  \citet{pietrukowicz03} identified eight more variables and 
both \citet{anderson03} and \citet{pietrukowicz05} announced one new variable.  
This brings the number of M22 variables to ninety, of which twenty-seven 
are RR Lyrae variables.

The sample of RR Lyrae stars in and around M22 is presented in Table~\ref{lcpars}.  
We estimate that the astrometry presented is accurate to better than 0.1~arcseconds 
and the photometry is accurate to $\sim$0.01 mag.  The columns contain 
(1) the name of the variable as given in the 2011 update of  M22 in the 
\citet{clement01} catalog, 
(2) the right ascension in hours, minutes and seconds (epoch J2000), 
(3) the declination in degrees, arcminutes and arcseconds, 
(4) the period in days, 
(5-8) the magnitude-weighted mean $U$, $B$, $V$, and $I$, respectively,
(9-12) the $U$-, $B$-, $V$- and $I$-amplitude, respectively,
(13) the type of variable,
(14) the projected radius (in arcmin) from the star to cluster center ($\rm R_{proj}$),
(15) the separation index (as defined in Stetson, Bruntt \& Grundahl 2003)
and 
(16) any comments.

\begin{figure}[htb]
\includegraphics[width=1\hsize]{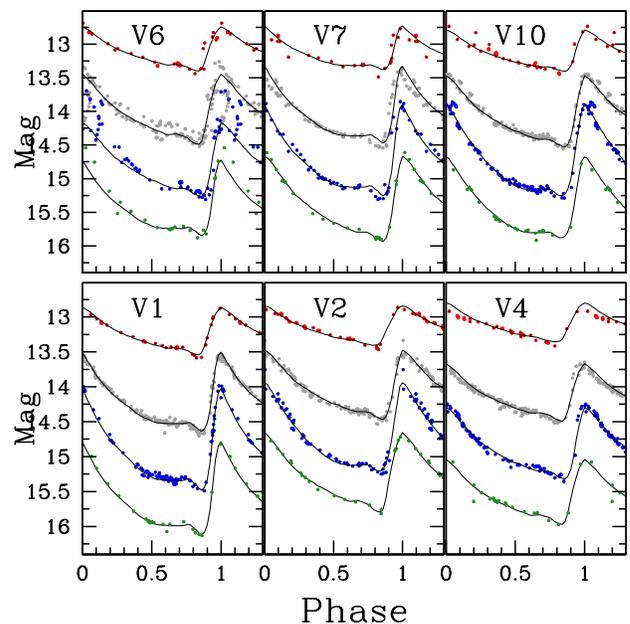}
\caption{Presentation of the phased $UBVI$ light curves of the fundamental
mode RR Lyrae variables in M22.
\label{lcrr0a}}
\figurenum{1}
\end{figure}

%
%
%
%
The light curves of the RR0 Lyrae variables are presented in Figure~\ref{lcrr0a}, and
the RR$1$ stars are shown in Figure~\ref{lcRR1a}.  The template-fitting routines 
from \citet{layden98} and \citet{layden00} were used to fit the data, but a Fourier
decomposition was used for their mean magnitude and amplitude parameters.
Our observations do not cover the RR Lyrae variables V22 and V27, as they lie 16.5 
and 20 arcminutes, respectively, from the cluster center.  These stars were observed 
as part of the ASAS survey, and their bright mean magnitudes ($\sim$1 mag brighter) 
suggest that both these stars are too bright to belong to the cluster.  Further evidence 
that these two stars belong to the field population comes from their amplitudes, which 
are small for their pulsation periods (see Table~\ref{lcpars}), suggesting they are 
more metal-rich than the rest of the M22 RR Lyrae stars.  Our observations also 
provide insufficient phase coverage for V3, but our photometry clearly suggests that 
this star is too faint to belong to the cluster.  It has a magnitude similar to the
majority of the RR Lyrae variables located in the Galactic bulge \citep{kunder08, pietrukowicz12}.
Similarly the RR Lyrae variable KT-12 is too faint to belong to M22, and also likely 
belongs to the bulge population; our phase coverage for this star allows robust 
$UBVI$ magnitudes to be determined, which are presented in Table~\ref{lcpars}. 

We have discovered four new low amplitude RR Lyrae stars, presented in Figure~\ref{lcRR1nv}.
These stars are all located $\sim$6 arcminutes from the center of the cluster; 
there are only 3 other M22 RR Lyrae variables at comparably large distances, 
the other M22 RR Lyrae variables lying at a mean distance of 2 arcminutes from the center.
The specific frequency of RR Lyrae stars in M22 is $S_{RR}$=10.4 where 
$S_{RR} = N_{RR} \times 10^{0.4(7.5+M_V)}$, and $M_V$=$-$8.50 \citep[][2010 update]{harris96} 
is the cluster's integrated absolute magnitude in $V$.  

The mean periods for the 10 RR0 Lyrae stars and 16 RR$1$ stars are 
$\langle P \rangle_{RR0}=$0.66$\pm$0.02 d and $\langle P \rangle_{RR1}=$0.33$\pm$0.01 d, 
respectively.  The ratio of RR$1$ to total RR Lyrae stars, $N_1/N_{RR}$, is 0.61.
As Figure~\ref{Nratio} shows, the mean periods of the RR0 Lyrae stars are similar to those
found in the typical Oosterhoff II (OoII) clusters, as is the large $N_1/N_{RR}$ ratio
\citet[e.g.,][]{catelan09}.


%
%
\begin{figure}[htb]
\includegraphics[width=1\hsize]{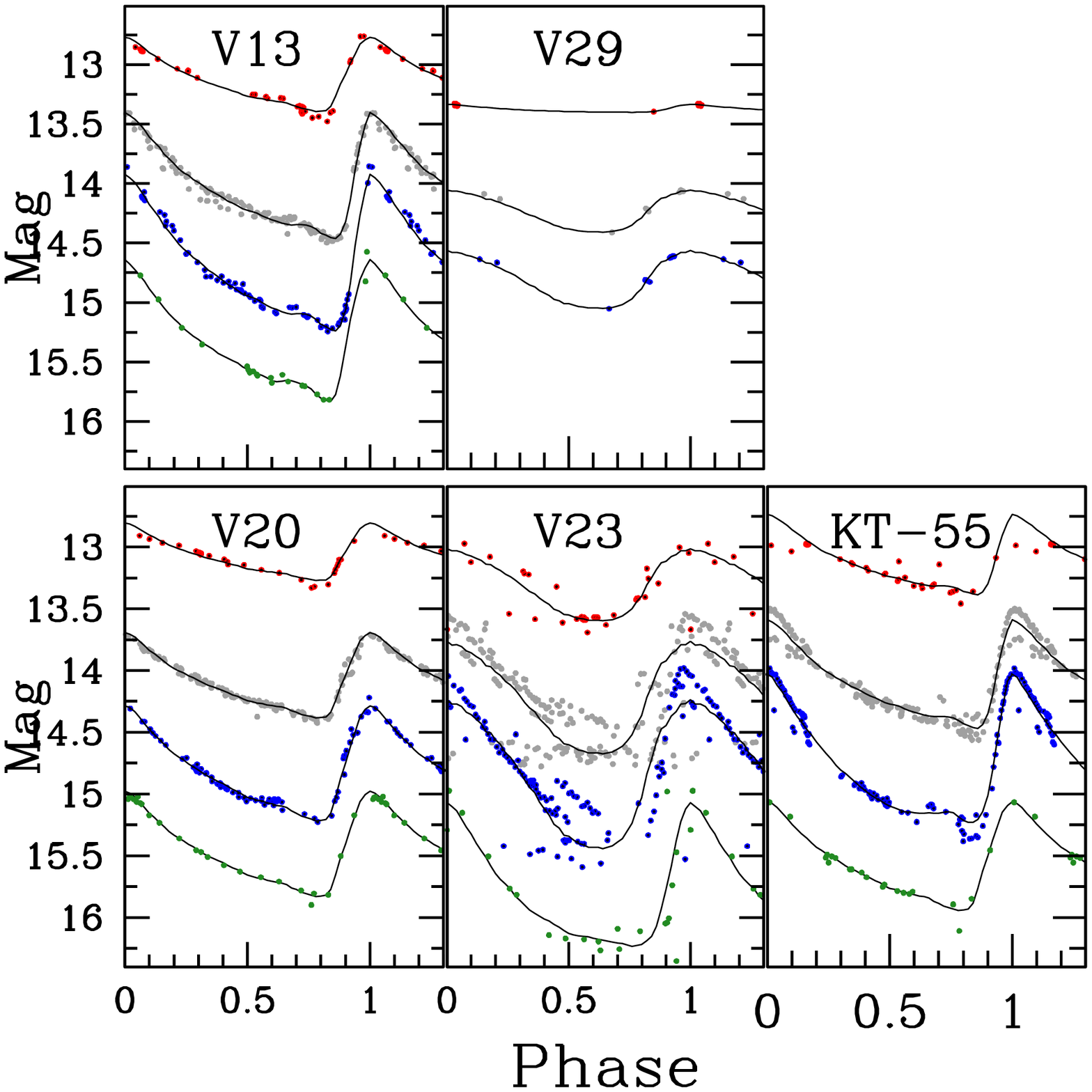}
\caption{ RR0 Lyrae light curves continued
\label{lcrr0b}}
\end{figure}
\begin{figure}[htb]
\includegraphics[width=1\hsize]{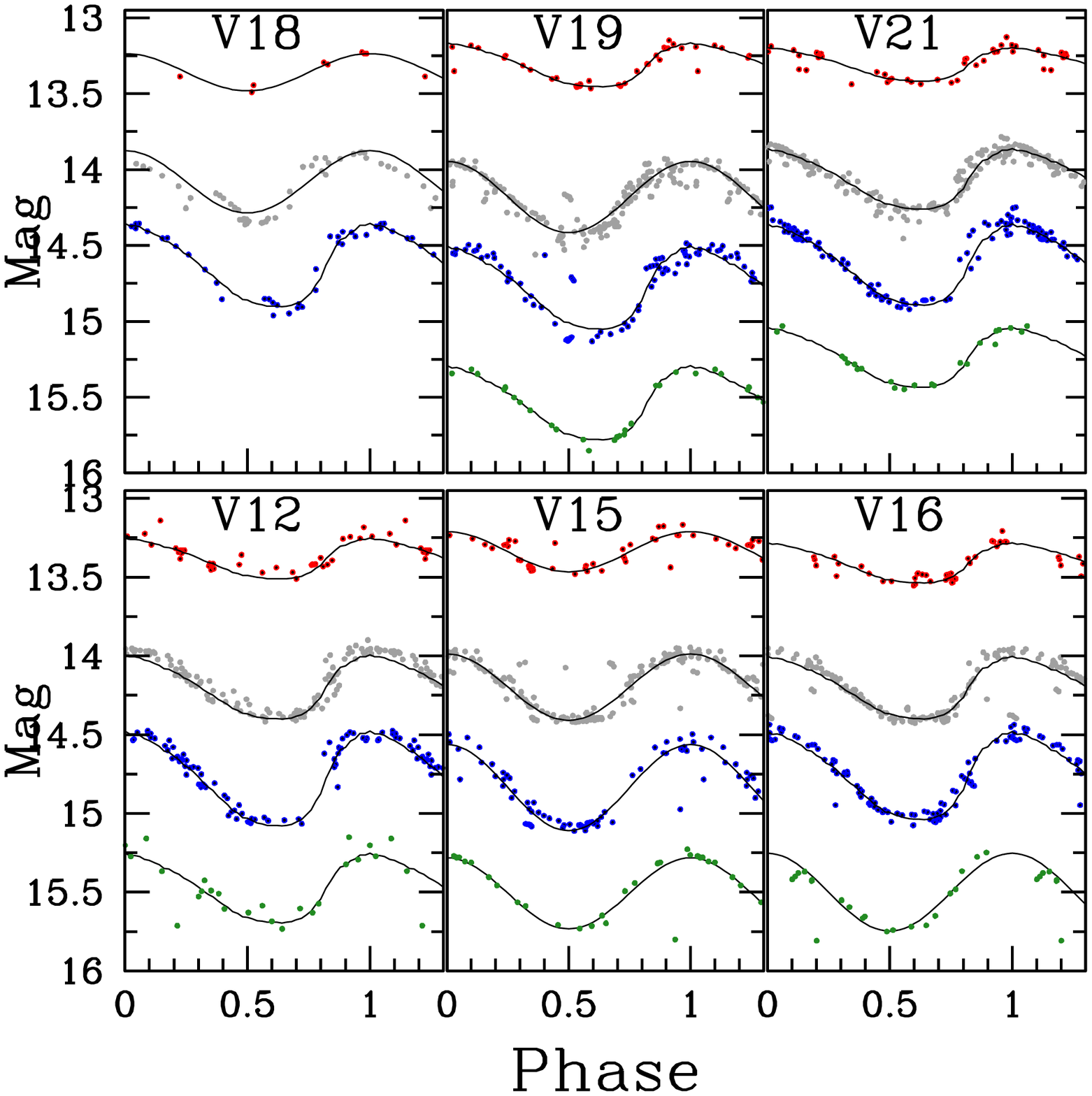}
\caption{ Presentation of the phased $UBVI$ light curves of the first overtone
RR Lyrae variables in M22.
\label{lcRR1a}}
\figurenum{1}
\end{figure}
\begin{figure}[htb]
\includegraphics[width=1\hsize]{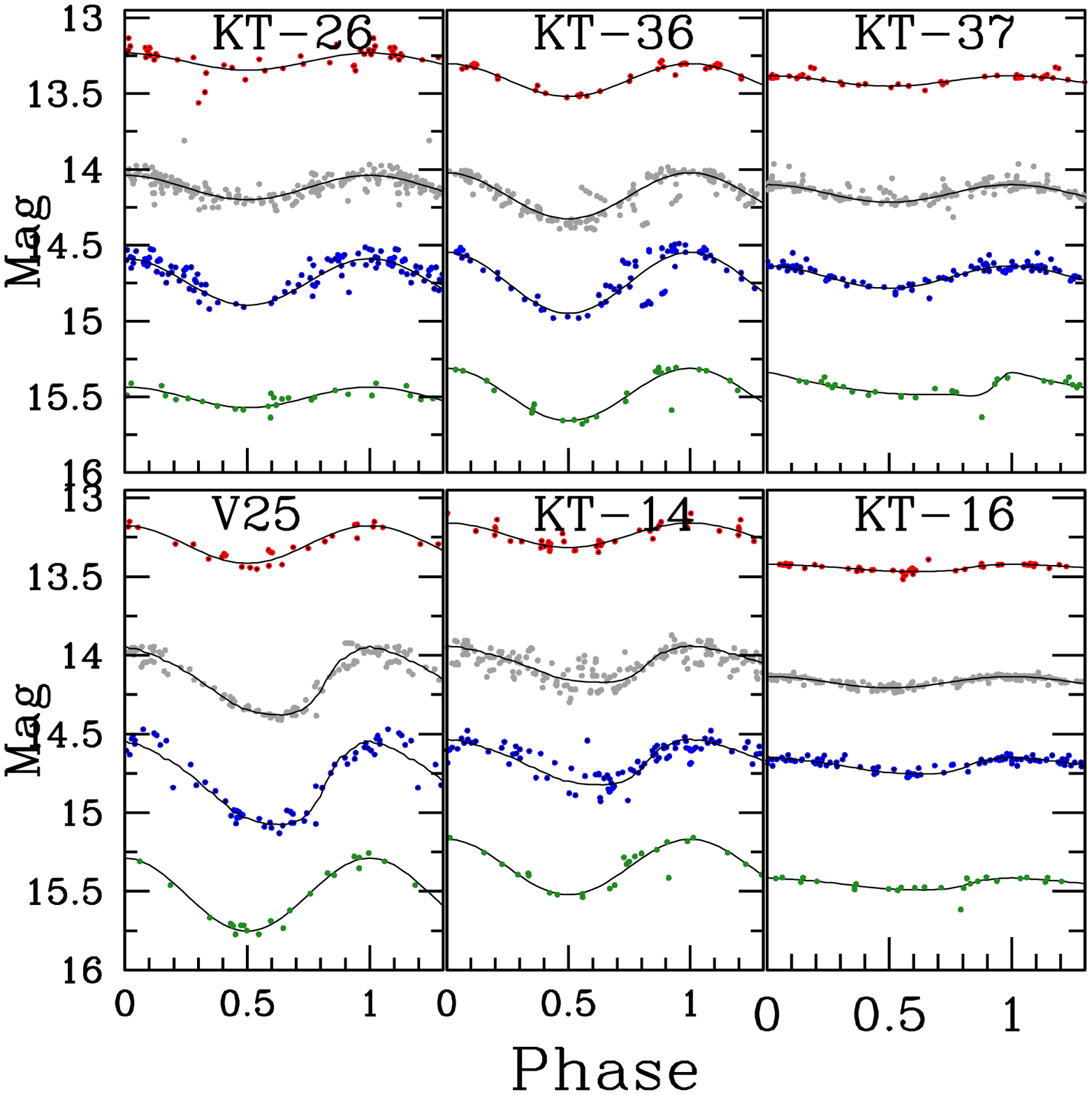}
\caption{ RR$1$ light curves continued
\label{lcRR1b}}
\end{figure}
\begin{figure}[htb]
\includegraphics[width=1\hsize]{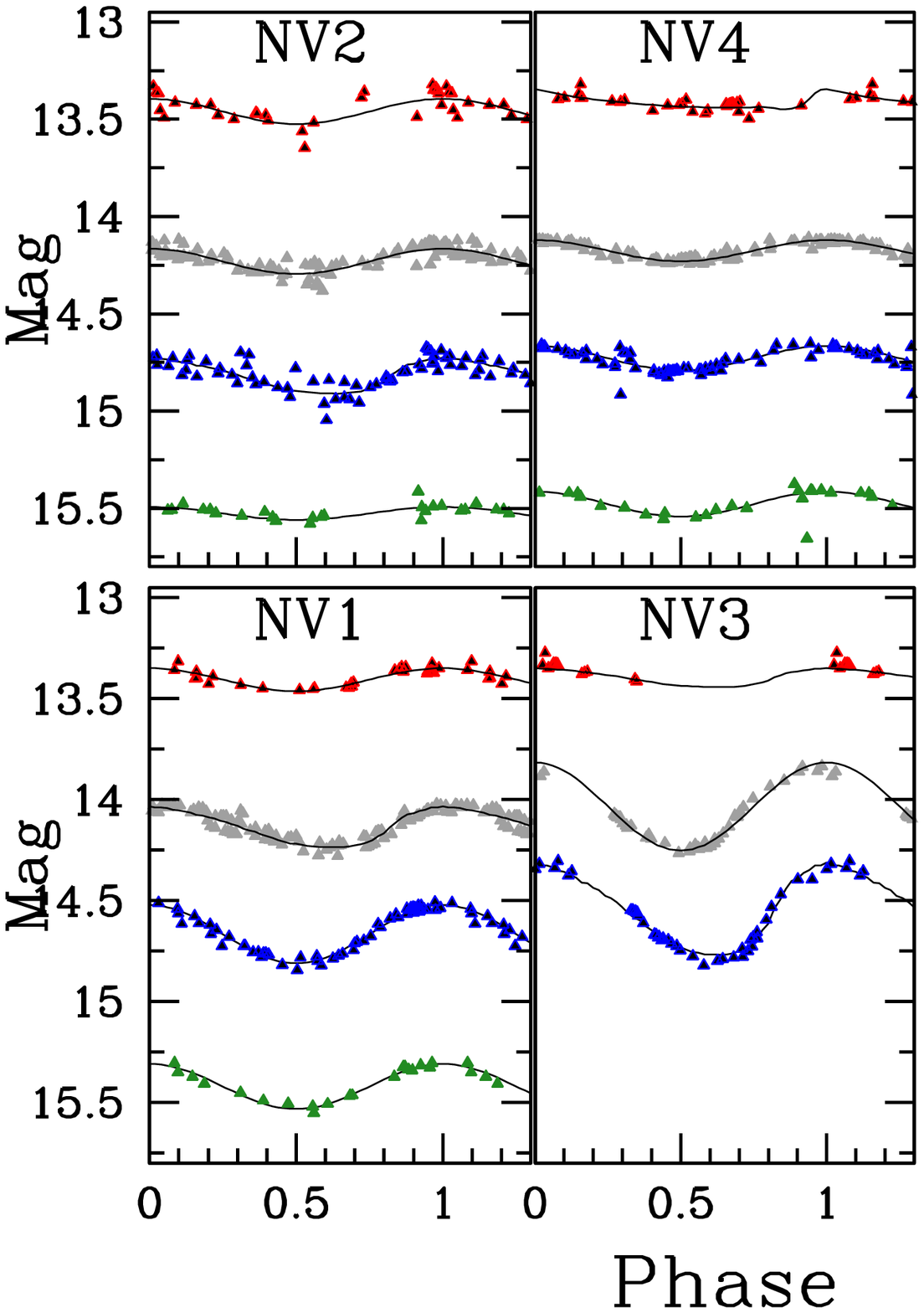}
\caption{Presentation of the phased $UBVI$ light curves of the newly discovered
RR Lyrae variables in M22.
\label{lcRR1nv}}
\end{figure}
\begin{figure}[htb]
\includegraphics[width=1\hsize]{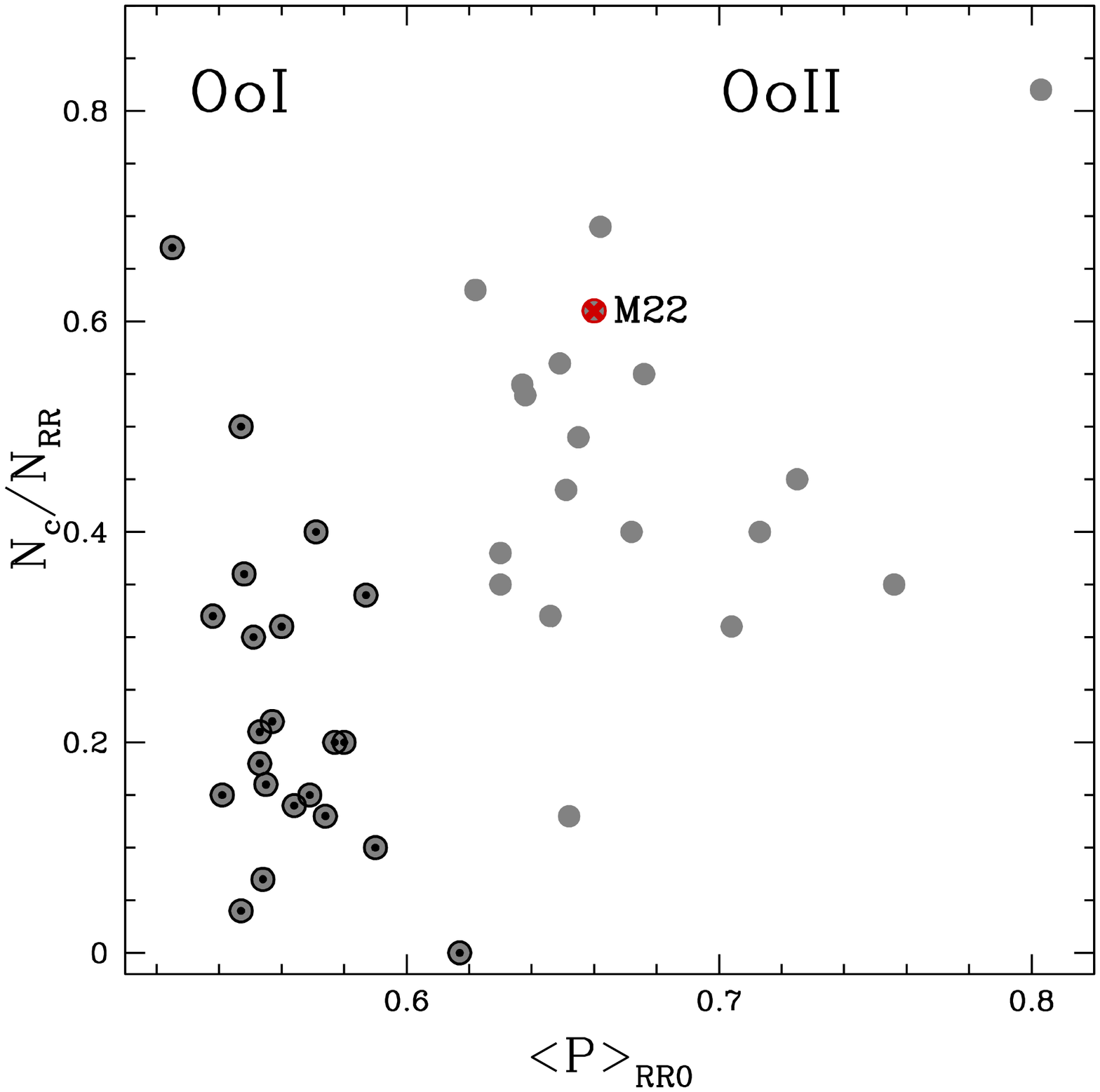}
\caption{
The ratio of RR$1$ to total RR Lyrae stars, $N_1/N_{RR}$, shown as a function of average RR0
period, based on the compilation presented in Catelan 2009.  The properties of M22
derived here indicated this cluster is an OoII-type GC.
\label{Nratio}}
\end{figure}

A significant fraction of RR Lyrae variables exhibit the Blazhko effect (hereafter 
also referred to as ``Blazhkocity"), which can be described as
long-term, sometimes periodic but often irregular amplitude and/or phase 
modulations \citep[e.g.,][]{blazhko07,klepikova56,benko10, buchler11,leborgne12}.  
This effect is not understood theoretically.  
The only systematic survey carried out to determine the frequency of light curve 
modulation of RR0 Lyars stars is the Konkoly Blazhko Survey \citep[KBS,][]{jurcsik09}.
KBS1 consists of 30 field RR0 Lyrae stars with $P_{RR0}$$<$0.50 d, 
and it was found that $\sim$47\% exhibit Blazhkocity \citep{jurcsik09}.  
Similarly, from 124 field RR0 stars with periods ranging from 0.55--0.60 d, KBS2 reports a
$\sim$43\% incidence of stars with Blazhkocity \citep{sodor12}.  
The case for the RR$1$ stars is more unclear.  Blazhkocity is usually not seen in 
RR$1$ variables \citep[it is thought that the Blazhko effect occurs in $\sim$5\% of RR$1$ 
variables;][]{moskalik03, kolenberg11} although recent studies of the GCs NGC$\,$2808 
and M53 have revealed RR Lyrae populations with large ($>$ 50\%) RR$1$ Blazhko 
percentages \citep{arellano12, kunder13b}.  

The RR0 stars in M22 with light curve modulation include V6, V23 and KT-55 (see Figure~\ref{lcrr0a}).  
However, upon closer examination, the 
scatter in the light curve of V6 appears to be caused by various phase jumps, with no 
obvious cyclic variation of the pulsation amplitude such as seen in stars exhibiting Blazhkocity.
V23 also has a light curve more indicative of a star with a rapidly changing
or erratic period, as opposed to a star with Blazhkocity.  KT-55, on the other hand, 
may be exhibiting Blazhkocity, particularly as amplitude modulation at maximum light is seen
in the $V$ light curve (where there are more observations.)  Assuming that only KT-55 exhibits 
Blazhkocity, this suggests a relatively small  ($\sim$10\%) incidence of Blazhko RR0s in M22.

However, 
%
%
although our observations span many years, the most useful set of observations were
taken over three nights in 2012 and such irregular spacing
is not sufficient to derive either Blazhko periods or secondary periodicities with any confidence.  
Without Blazhko periods, it is difficult
to know for certain whether the scatter in the light curves is indeed due to the Blazhko effect 
or whether the light curve scatter is due to other phenomena
such as unstable periods, 
large period-change rates \citep[e.g.,][]{leborgne07, kunder11}, secondary 
periods \citep[e.g.,][]{fitch76, peniche89, moskalik03}, 
period doubling \citep[e.g.,][]{szabo10}, and/or photometric anomalies such as blending compounded by variable seeing
conditions.
We have searched the stars for secondary periods, and find no firm evidence of secondary periods in 
the sample.  
However, it is worth noting that double-mode
RR Lyrae stars usually reside in metal-poor OoII-type clusters \citep[e.g.,][]{clement91}
so the environment of M22 could be receptive to such pulsators. 


We have also searched for trends between light curve scatter
and stellar brightness to see whether any faint neighboring stars 
were contaminating the stellar PSF profile (as blending would make the RRL brighter, while
blending combined with variable seeing could increase the light curve scatter).
No systematic trend in perceived stellar brightness
as a function of radial distance is seen, and no trends in stellar brightness 
as a function of Blazhkocity are found either.  However, it is worth mentioning
that many of the stars with light curve scatter are located closer to the center of the cluster.
We have also searched for trends in a stars' separation index, where the separation index is
defined by Stetson, Bruntt \& Grundahl (2003).  The larger the separation index,
the less a given star is contaminated by its known neighbors.  All of the RR Lyrae 
variables have separation indices greater than zero, although there are
two stars (one exhibiting light curve scatter and one that does not) with a separation
index less than or equal to three.  A separation index of 3.0 indicates that the light of
a star is contaminated by its {\it known\/} neighbors at a 6\% level; since the known neighbors
are accounted for in the PSF fits, the residual contamination by these stars should be at least
an order of magnitude less.  In summary, we stress that we have no specific evidence 
that the light curve scatter seen is caused by blending.
But we do remain alert to the fact that PSF fitting in crowded fields observed
under different 
seeing conditions is challenging. It is also possible that incorrect cycle counts spanning
long intervals without observations can lead to inaccurate periods, but we stress that the
periods we have derived are statistically superior to alternative
periods requiring different numbers of cycles over the time span that we have covered.


\subsection{RR Lyrae Distance}
The mean apparent magnitude for the RR Lyrae variables is 
$\rm \langle m_{V,RR}\rangle$ = 14.15 $\pm$ 0.02 mag, where the confidence interval is the 
standard error of the mean.  This is in excellent agreement with the \citet{harris96} 
value of $\rm m_{V,HB}$ = 14.15, which lists the mean $V$ magnitude of the 
horizontal branch, $V_{HB}$.  It also agrees well with the \citet{monaco04} estimate 
of  $\rm m_{V,HB}$ = 14.17 $\pm$ 0.25 mag,
found from averaging the $V$ magnitudes of the RR Lyrae stars observed at random phases.

The absolute magnitudes of the RR0 variables can be estimated using:
\begin{equation}
M_V = 0.23 {\rm [Fe/H]_{CG97}} + 0.931
\end{equation}
from \citet{catelancortes08}, where $\rm [Fe/H]_{CG97}$
is the metallicity in the \citet{carretta97} scale.
Adopting \feh=$-$1.75 dex on the \citet{kraft03} scale \citep{marino09, dacosta09}, 
$\rm [Fe/H]_{CG97}$=$-$1.55 dex, and therefore the absolute magnitudes
of the RR Lyrae variables in M22 are $M_V$=0.57$\pm$0.13 mag. 
Similarly, using a quadratic relation between 
RR Lyrae absolute magnitude and metallicity from Bono, Caputo, \& di Criscienzo (2007):
\begin{equation}
M_V = 0.08{\rm [Fe/H]^2} + 0.50{\rm [Fe/H]} + 1.19
\end{equation}
the RR Lyrae absolute magnitudes are $M_V$=0.56$\pm$0.08.
A brighter $M_V$=0.40$\pm$0.07 is found when using the \citet{benedict11}
$M_V$--$\rm [Fe/H]$ relation, however, suggesting that the level of agreement between 
independent $M_V$ measurements is as large as $\sim$0.2 mag.


The level of evolution off the ZAHB can 
also affect an RR Lyrae star's absolute magnitude in $V$ by $\sim$0.08 
mag \citep[e.g.,][]{sandage90, clementini03},
as can an RR Lyrae star's helium content, alpha-element abundance and CNO content.
The metallicity bimodality in M22 of ${\rm \Delta[Fe/H]}$$\sim$0.14 
(see the Introduction), corresponds to variation of 0.04 mag in $M_V$.

It is worth noting that the \citet{catelancortes08} absolute magnitude relationship assumes a 
helium abundance of $Y$=0.23, an $\rm [\alpha/Fe]$= +0.31, and no CNO enhancement.  
The \citet{bono07} relation is based on synthetic HB simulations and a large sample of Galactic 
globular clusters; therefore it 
assumes a He, $\rm [\alpha/Fe]$ and CNO typical of MW GCs.  If the He, $\rm [\alpha/Fe]$ and 
CNO of the M22 RR Lyrae stars is abnormal, the RR Lyrae absolute magnitudes will be slightly affected.

Using the above determined $M_V$ of $M_V$=0.57$\pm$0.12, the apparent distance 
modulus of M22, based upon the RR Lyrae variables, is $(m-M)_{V,\rm RRL}$=13.58$\pm$0.13.
Using  $E(\hbox{\it B--V\/})$=0.36 mag (see below, \S3.5), we obtain a
true distance modulus $(m-M)_{0,\rm RRL}$=12.46$\pm$0.13, which implies a geometric
distance of 3.1 $\pm$ 0.2 kpc.  
The \citet{benedict11} absolute magnitue of $M_V$=0.40$\pm$0.07 results in $(m-M)_{V,\rm RRL}$=13.75$\pm$0.11.
and a true distance modulus $(m-M)_{0,\rm RRL}$=12.67$\pm$0.12 or 3.4 $\pm$ 0.2 kpc. 

RR Lyrae period-luminosity PL relations in $IJHK$ are powerful in that they are not
as sensitive to evolutionary effects as the bluer passbands, 
leading to the presence of tighter absolute magnitude relations \citep[e.g.,][]{catelan04}.  
The reddening is also not as severe in these passbands.
Using the \citet{catelan04} relation of
\begin{equation}
M_I = 0.471 - 1.132 {\rm log} P+ 0.205 {\rm log} Z
\end{equation}
and Z=0.0006 ($\feh$=$-$1.77, $Y$=0.23, $\alpha/Fe$=+0.35, no CNO enhancement),
and using  $E(\hbox{\it B--V\/})$=0.36 mag, a
true distance modulus $(m-M)_{0,\rm RRL}$=12.50$\pm$0.10 is found.
This translates to a geometric distance of 3.2 $\pm$ 0.2 kpc, and is in excellent agreement
with the distances determined from the $M_V$--\feh relations above.

\subsection{The RR Lyrae Color-Magnitude Diagram}
\begin{figure}[htb]  
\includegraphics[width=9cm]{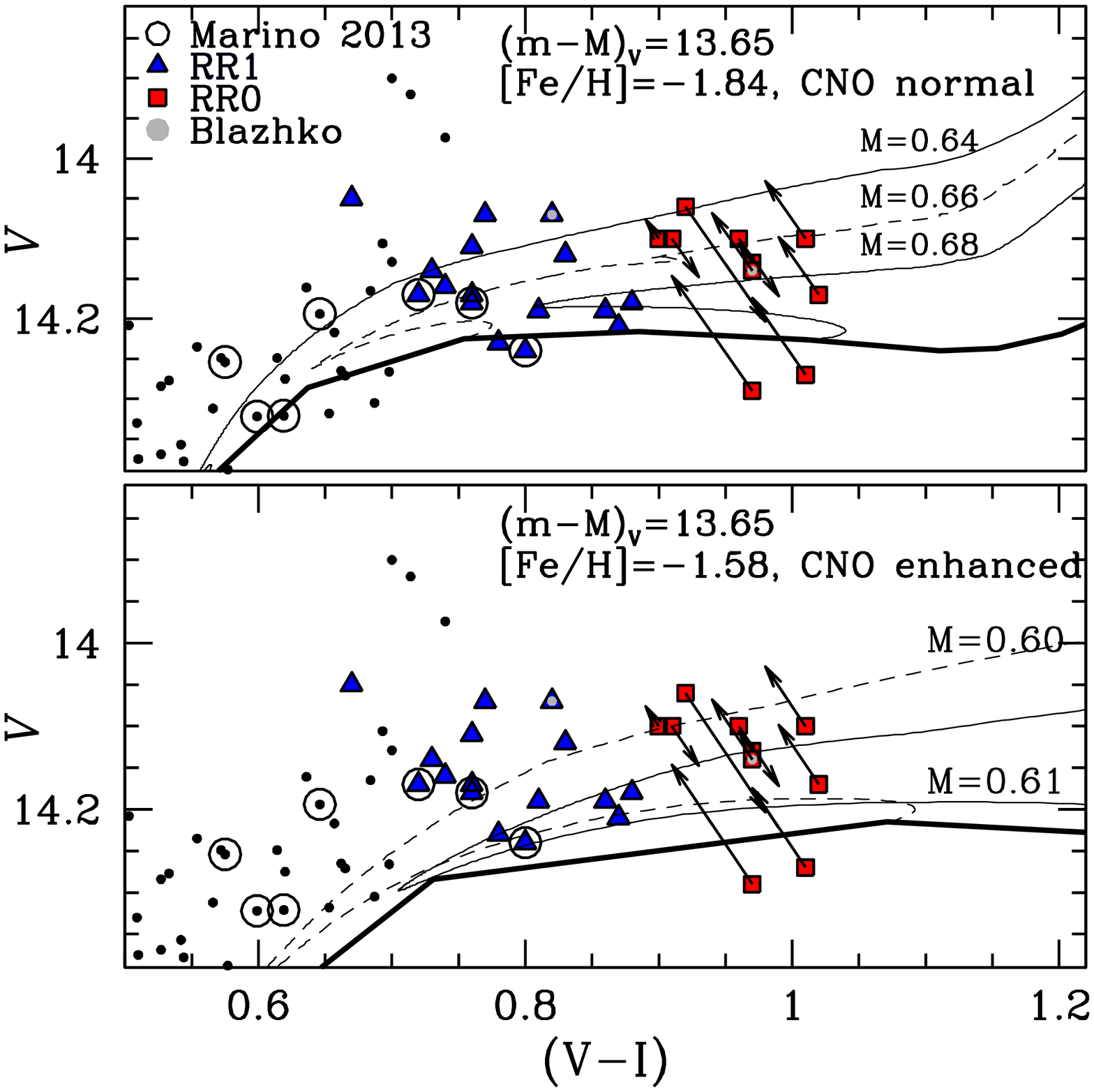}
\caption{The M22 Instability strip in $V$ versus \hbox{\it V--I\/}.  
Stars near the M22 instability strip are marked with points as defined in the key.  Curves are BaSTI models 
of the ZAHB (bold) and evolutionary tracks for different masses ($M$) for a metal-poor, CNO-normal case (top) 
and a metal-rich, CNO-enhanced case (bottom).  Arrows indicate the differential reddening corrections for 
RR0 stars described in \S3.3.
\label{cmdhb}}
\end{figure}

Figure~\ref{cmdhb} gives an expanded view of the HB.   
As suggested earlier and demonstrated in \S3.5, there is significant differential reddening across the face of 
M22 that increases the scatter in Figure~\ref{cmdhb}.  The RR0 stars provide a means to quantify and correct for 
this effect.  The arrows in Figure~\ref{cmdhb} show the originally observed positions (squares) and the positions 
corrected for differential reddening (arrow heads) based on the mean reddening of ${\rm E(\hbox{\it V--I\/})=0.46}$
and the stars' colors in Table~5.  Thus, the arrow heads should be compared to the evolutionary model tracks; 
this indicates that the RR0 in M22 are likely lower mass stars evolved from BHB rather than higher mass stars living 
on the red end of the ZAHB.  The arrows also give a sense for how much the RR1 and non-variable BHB stars in 
Figure~\ref{cmdhb} are affected by differential reddening.

%
%

As mentioned in the Introduction, 
the stars in M22 can be roughly divided into two metallicity groups -- the $s$-rich and 
the $s$-poor stars \citep[e.g.,][]{marino11, alvesbrito12}, with 
${\rm [Fe/H]_{s-rich}=-1.68\pm0.02}$ and 
${\rm [Fe/H]_{s-poor}=-1.82\pm0.02}$ \citep{marino09}.
The appropriate theoretical models from the BaSTI archive \citep{pietrinferni04, 
pietrinferni06, pietrinferni09} are overplotted in Figure~\ref{cmdhb}
to illustrate how the RR Lyrae stars in M22 are matched 
by theoretical evolutionary predictions.
In particular, the Zero Age Horizontal Branch (ZAHB) loci corresponding to different 
assumptions about the chemical mixtures are shown: a ZAHB locus for an $\alpha-$enhanced mixture with 
${\rm [Fe/H]=-1.84}$, and a ZAHB locus corresponding to a mixture where the CNO sum is 
enhanced with respect the standard $\alpha-$enhanced mixture and ${\rm [Fe/H]=-1.58}$. 
For the CNO-enhanced ZAHB locus, the BaSTI models available account for a CNO-sum enhancement 
of approximately ${\rm [C+N+O/Fe]\approx 0.3}$~dex.  However, as a smaller CNO-enhancement 
of $\sim0.13$~dex is found from the spectroscopic measurements  of M22, a 
correction to both magnitude and color index of the models
is made to account for this difference\footnote{This 
correction has been estimated by comparing the model predictions for various 
levels of CNO-enhancement discussed in \cite{pietrinferni09}.}.  
For the standard $\alpha-$enhanced stellar locus, a shift of +0.05 mag in $V$ to the BaSTI ZAHB 
is applied to account for the updated conductive opacities provided by \citet{cassisi07}.
This shift is important to obtain a self-consistent comparison with the CNO-enhanced ZAHB,
which has been computed accounting for the updated conductive opacities \citep[see]
[for a detailed analysis of this issue]{cassisi07}.

We adopt a reddening ${\rm E(B-V)=0.36}$ and a distance modulus of ${\rm (m-M)_V=13.65}$~mag 
for the theoretical comparison in Figure~\ref{cmdhb}.
This distance modulus is within the range of the distances obtained from the RR Lyrae variables discussed above in \S3.2
and was chosen because it allows the lower envelope of the observed HB star distribution
for the standard $\alpha-$enhanced ZAHB to match the BASTI models.  

In each panel of Figure~\ref{cmdhb}, 
the evolutionary tracks of some selected HB stellar models are also shown. 
Figure~\ref{cmdhb} therefore shows the comparison between observational data and 
theoretical models.  Because the theoretical models in both panels are located
at approximately the same position in the CMD, it is difficult to trace the origin of the 
RR Lyrae population -- that is, to assess whether they belong to the first generation 
(the canonical $\alpha-$enhanced one) or to the second generation (the more metal-rich, 
CNO-enhanced one).  However, it is worth noting that, that if one assumes that the RR 
stars are (mainly) associated with the first 
generation, then their location in the CMD is well matched by the evolutionary pattern
of HB models during the main core He-burning stage.  In particular, BaSTI models also 
indicate that in such a scenario, most of the M22 RR Lyrae stars 
have a rather small mass range of $\sim$0.66 -- 0.68$M_\sun$. 
We note that it has been shown spectroscopically that 
three (blue) RR1 variables in M22 are metal-poor with a normal CNO content \citep{marino13}
and therefore likely belong to the first generation.  

On the other hand, if one assumes that the RR Lyrae stars belong to the second generation, 
the theoretical prediction is that these variables are brighter than the ZAHB, and that
their location (at least of the majority of them) could be (mainly) explained as due to 
stars crossing the instability strip during their off-ZAHB evolution.  In this scenario, their 
predicted masses would be $\sim$0.60 -- 0.61$M_\sun$.

It is obvious that spectroscopic investigations of the chemical composition of the M22 RR Lyrae stars, 
and in particular the measurement of O and Na abundances, would 
help in discriminating between these scenarios.  However, due to the large difference in the 
evolutionary rates between HB stars experiencing their major core He-burning stage
and those crossing the RR Lyrae when moving from the blue side of the HB distribution 
towards the Asymptotic Giant Branch, some useful hint for disentangling the evolutionary 
origin of these variable stars could be obtained by the analysis of the secular pulsation 
period changes \citep[however see e.g.,][for the difficulties in associating period change rates
to the evolutionary status of a star]{leborgne07, kunder11}.  

%
%
%

\subsection{Period-Amplitude Diagram}
\begin{figure}[htb]  
\includegraphics[width=1\hsize]{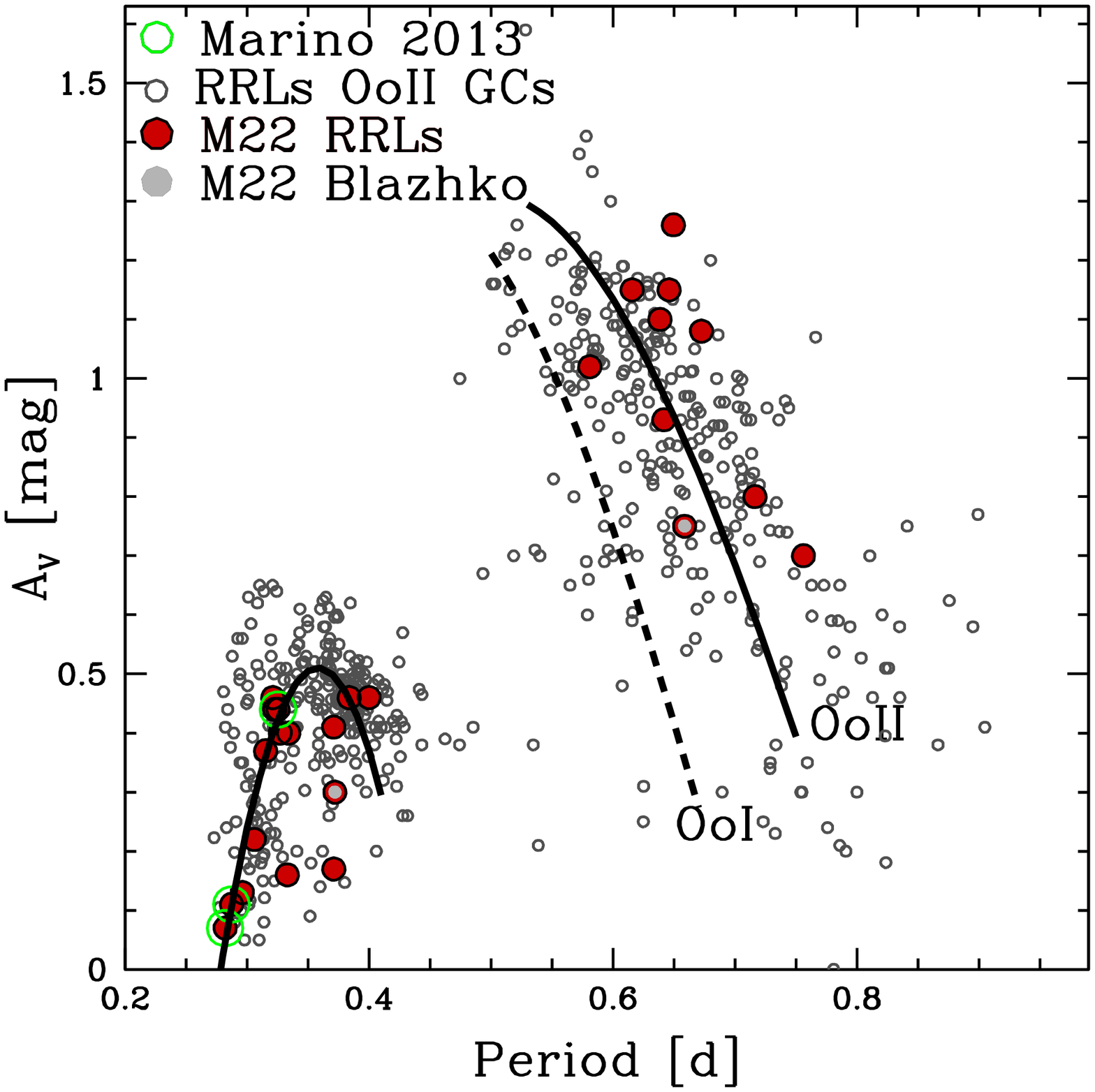}
\caption{Position of the M22 RR Lyrae stars on the period-$A_V$ diagram.  
The stars exhibiting Blazhkocity are represented by gray circles.  The dashed line
is typical for RR0 variables in OoI clusters and the solid line is for OoII clusters, 
according to \citet{cacciari05}.  The periods and $V$-band amplitudes for 
RR Lyrae stars in 14 OoII-type GCs are also shown by small circles (see text for details).
\label{PA}}
\end{figure}

The period-amplitude (PA) diagram for the RR Lyrae stars is shown in Figure~\ref{PA},
where $A_V$ is the amplitude in the $V$-band.
Also shown are the RR Lyrae stars in 14 OoII-type GCs (NGC$\,$4590, NGC$\,$5053, 
NGC$\,$7078, $\omega$~Cen, NGC$\,$5466, NGC$\,$6426, NGC$\,$6333, NGC$\,$6341, 
NGC$\,$5024, NGC$\,$7089, NGC$\,$2419, NGC$\,$5286, NGC$\,$6101 and NGC$\,$1904).
The period-$A_V$ relation for OoI and OoII clusters is overplotted, where the 
fundamental-mode and first-overtone PA relations are derived from the M3 RRL \citep{cacciari05}.

The M22 RR0 variables occupy the area of the PA diagram where OoII-type variables tend to 
lie \citep[e.g.,][]{clement99}.  Unlike RR0 variables, RR$1$ variables are not expected to 
follow a roughly linear sequence in the PA diagram \citep[e.g.,][]{bono97}, and this is clearly
seen within the RR$1$ population of M22.  The ``hairpin"
shape predicted for the RR$1$ variables in the theoretical PA diagram is evident,
albeit with significant outliers.
The approximate relation

\begin{equation}
A_V = -9.75 + 57.3  P - 80  P^2
\end{equation}

\noindent
is derived from the M22 RR1 variables.  
Blazhkocity can affect the position of a star on the PA diagram 
\citep[e.g.,][]{cacciari05} and as discussed above, some of the RR Lyrae stars in M22 
may exhibit signs of this
phenomenon.  The amount of the amplitude change due to the Blazhko effect is
no more than $\sim$0.1 mag in $V$ (see Figure~\ref{lcRR1a}), which does not significantly
affect the location of the star in the PA diagram.

\begin{figure}[htb]  
\includegraphics[width=9cm]{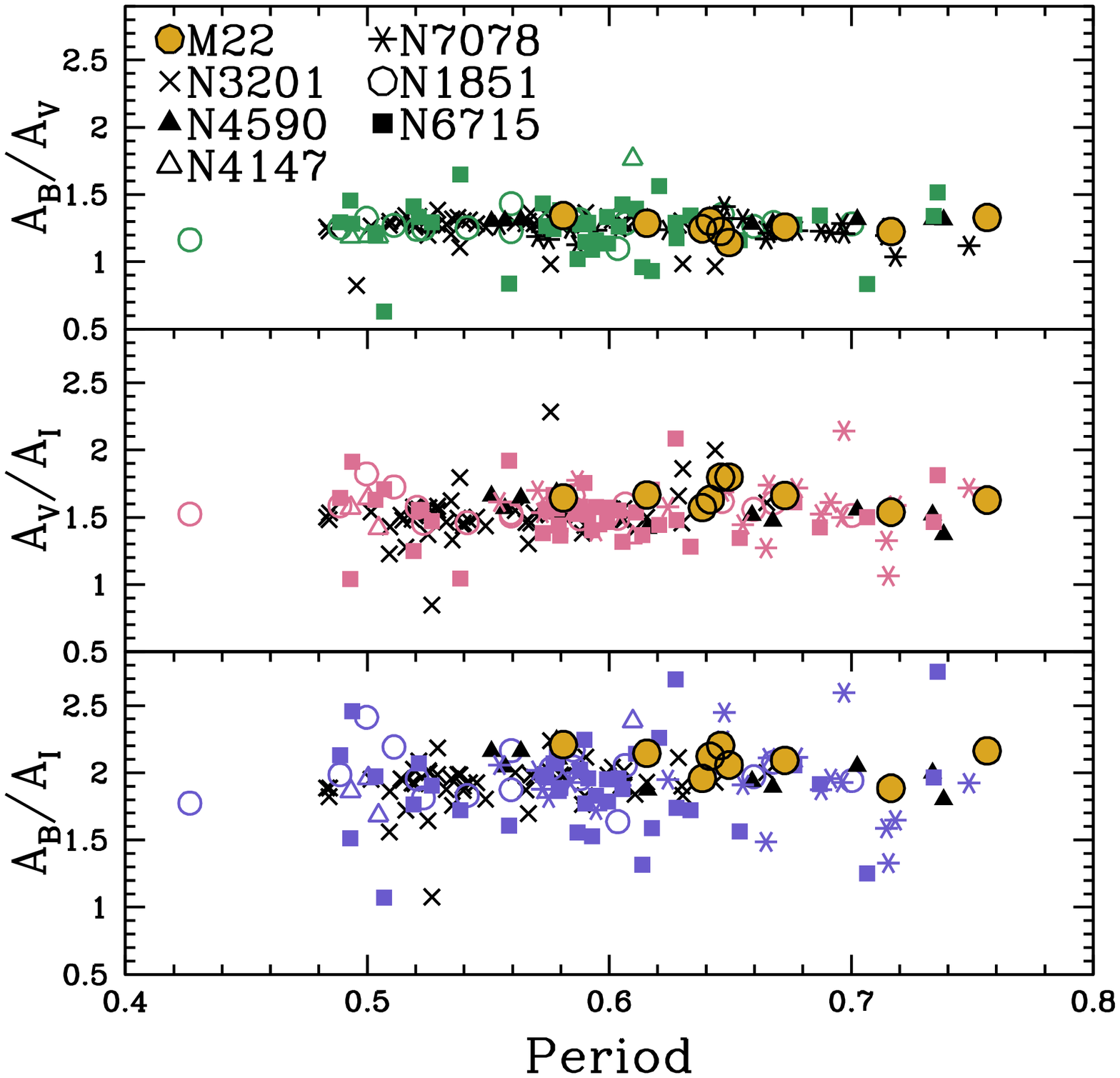}
\caption{The RR Lyrae amplitude ratios as a function of period for seven Milky Way globular clusters
(see Table~\ref{ampratio}).
\label{AmpRatioPer}}
\end{figure}

Amplitude ratios are often used both to identify variables that might have their 
photometry compromised by faint companions and to classify variable stars, 
especially the low amplitude RR Lyrae variables.  In light of this, the ratios of the 
RR0 and RR$1$ amplitudes in the different bands are given in Table~\ref{ampratio}
and Table~\ref{ampratio1}.  We have
excluded the RR0 star KT-55 in the amplitude ratio determination, as it has an uncertain
$A_I$ (see Table~\ref{lcpars}).  Similarly, the RR$1$ stars NV2, NV3 and V18 
are not included in the amplitude ratios due to no/uncertain $A_B$ and $A_I$s.
The amplitude
ratios of the RR Lyrae stars in few Galactic GCs are also listed for comparison purposes
and plotted as a function of their period in Figure~\ref{AmpRatioPer}.
There is a scatter of $\sim$0.2 mag for $A_B/A_I$ and $A_V/A_I$, but a much smaller
scatter of $\sim$0.05 mag, for $A_B/A_V$.  Moreover, \citet{kuehn11} find 
$A_B/A_V$ = 1.28 $\pm$ 0.02 from 26 RR Lyrae variables in the OoI-type LMC GC 
NGC$\,$1466 (see their Figure~9),
and \citet{kinman10} find $A_B/A_V$ = 1.27 from 12 field RR Lyrae stars in
the Northern Sky Variability Survey (NSVS) data.  These results are in excellent agreement with 
the $A_B/A_V$ amplitude ratios obtained here and in other Galactic GCs.  
This may suggest that Galactic field and cluster RR Lyrae and cluster LMC
RR Lyrae stars have $A_B/A_V$ RR Lyrae amplitude 
ratios that are independent of their host system.

\begin{table}
\begin{scriptsize}
\centering
\caption{Empirical amplitude ratios of the RR0 Lyrae variables}
\label{ampratio}
\begin{tabular}{p{0.4in}p{0.4in}p{0.4in}p{0.4in}p{0.4in}p{0.15in}p{0.35in}} \\ \hline
cluster & $A_U/A_I$ & $A_B/A_I$ & $A_V/A_I$ & $A_B/A_V$ & num stars & source \\ 
\hline
M22 & 1.89$\pm$0.05 & 2.09$\pm$0.04 & 1.66$\pm$0.03 & 1.26$\pm$0.02 & 9 & this work \\
NGC$\,$3201 & -- & 1.88$\pm$0.04 & 1.49$\pm$0.03 & 1.26$\pm$0.02 & 56 & (1) \\
NGC$\,$1851 & -- & 2.00$\pm$0.04 & 1.57$\pm$0.02 & 1.27$\pm$0.02 & 19 & (2) \\
NGC$\,$4147 & -- & 2.19$\pm$0.24 & 1.74$\pm$0.25 & 1.29$\pm$0.12 & 5 & (3) \\
NGC$\,$4590 & -- & 2.03$\pm$0.04 & 1.55$\pm$0.03 & 1.31$\pm$0.01 & 12 & (4) \\
NGC$\,$7078 & -- & 1.93$\pm$0.05 & 1.58$\pm$0.04 & 1.23$\pm$0.01 & 25 & (5) \\
NGC$\,$6715 & -- & 1.89$\pm$0.05 & 1.53$\pm$0.03 & 1.25$\pm$0.03 & 40 & (6) \\

\hline
\\
\end{tabular}
\end{scriptsize}
\end{table}
\begin{table}
\begin{scriptsize}
\centering
\caption{Empirical amplitude ratios of the RR$1$ variables}
\label{ampratio1}
\begin{tabular}{p{0.4in}p{0.4in}p{0.4in}p{0.4in}p{0.4in}p{0.15in}p{0.35in}} \\ \hline
cluster & $A_U/A_I$ & $A_B/A_I$ & $A_V/A_I$ & $A_B/A_V$ & num stars & source \\ 
\hline
M22 & 1.83$\pm$0.10 & 2.17$\pm$0.07 & 1.61$\pm$0.07 & 1.37$\pm$0.06 & 13 & this work \\
NGC$\,$3201 & -- & 1.97$\pm$0.02 & 2.05$\pm$0.07 & 1.02$\pm$0.04 & 4 &  (1) \\
NGC$\,$1851 & -- & 2.04$\pm$0.08 & 1.63$\pm$0.05 & 1.25$\pm$0.02 & 8 &  (2) \\
NGC$\,$4147 & -- & 2.06$\pm$0.14 & 1.54$\pm$0.25 & 1.38$\pm$0.18 & 10 & (3) \\
NGC$\,$4590 & -- & 2.10$\pm$0.02 & 1.62$\pm$0.02 & 1.30$\pm$0.01 & 16 & (4) \\
NGC$\,$7078 & -- & 1.97$\pm$0.06 & 1.55$\pm$0.04 & 1.28$\pm$0.03 & 13 & (5) \\
NGC$\,$6715 & -- & 1.7$\pm$0.1 & 1.56$\pm$0.02 & 1.08$\pm$0.08 & 8 & (6) \\
\hline
\\
\end{tabular}
(1) \citet{layden03}; (2) \citet{walker98}; (3) \citet{stetson05}; (4) \citet{walker94};
(5) \citet{corwin08}; (6) \citet{sollima10}
\end{scriptsize}
\end{table}
\subsection{RR Lyrae Reddenings}
The interstellar extinction toward M22 is high 
\citep[$E(\hbox{\it B--V\/})$ = 0.34 mag; 2010 edition of][]{harris96} 
and variations of reddening are expected within the field of view of 
M22 \citep[e.g.,][]{monaco04, piotto12}.
Minimum-light colors of the RR0 Lyrae variables have been used as a tool for measuring 
the interstellar reddening toward RRLs \citep{sturch66, walker90, blanco92, mateo95, kunder10}.
\citet{sturch66} derived a formula to determine reddenings toward RR Lyrae stars using 
their $\hbox{\it (B--V)\/}$ colors at minimum light, which was then corrected to be on the 
GC scale by \citet{walker90}.  In particular, for RR0 variables, \citet{walker90} obtained:
\begin{equation}
E(\hbox{\it B--V\/}) = (\hbox{\it B--V\/})_{min} - 0.24\ P - 0.056\ {\rm [Fe/H]_{ZW84}} - 0.336
\label{sturchwalker}
\end{equation}
where $(\hbox{\it B--V\/})_{min}$ is the color at minimum light (phase between 0.5 and 0.8), $P$ is
the period in days and $\rm [Fe/H]_{ZW84}$ is the metallicity in the
\citet{zinn84} scale.
Similarly, using the $V$ and $I$ passbands, \citet{guldenschuh05}
find:
\begin{equation}
E(\hbox{\it V--I\/}) = (\hbox{\it V--I\/})_{min} - 0.58\pm0.02
\label{laydeng}
\end{equation}
and that this relation is largely independent of period, amplitude and metallicity.

The observed \hbox{\it B--V\/} and \hbox{\it V--I\/} color at minimum light is calculated 
using the best-fit light-curve templates and listed in Table~\ref{evmi}.  Comparing 
colors at minimum light derived from fitted template light curves, to minimum-light 
colors estimated directly from the observation, we find that the values agree usually 
to within 0.01 mag.  
The average $E(\hbox{\it B--V\/})$ from Equation~\ref{sturchwalker}, adopting
$\rm [Fe/H]_{ZW84}$=$-$1.73 dex \citep[which is \feh=$-$1.75 on the][scale]{kraft03}, 
gives $E(\hbox{\it B--V\/})$=0.38 $\pm$ 0.02 mag. 
The average $E(\hbox{\it V--I\/})$ from Equation~\ref{laydeng} 
is 0.47 $\pm$ 0.03 mag  
which translates to $E(\hbox{\it B--V\/})$ of 0.34 - 0.37 mag, depending
on the reddening law used. \footnote{Different values for the transformation 
from $E(\hbox{\it V--I\/})$ to $E(\hbox{\it B--V\/})$ include a 
$E(\hbox{\it V--I\/}) / E(\hbox{\it B--V\/})$ ratio of 
1.4 \citep{schlegel98, schlafly11}, 1.35 \citep{mccall04} and 1.28 \citep{cardelli89, cardelli92}.}  
%
%
%
%
%
Adopting an average $E(\hbox{\it V--I\/}) / E(\hbox{\it B--V\/})$ ratio of 1.35, 
$E(\hbox{\it B--V\/})$ is 0.35 $\pm$ 0.04 mag. 

The average of the $E(\hbox{\it B--V\/})$ values from Equation~\ref{sturchwalker} and
Equation~\ref{laydeng} is $E(\hbox{\it B--V\/})$=0.36 $\pm$ 0.03, in
excellent agreement with the value of $E(\hbox{\it B--V\/})$=0.38 $\pm$ 0.04 from
\citet{monaco04} and \citet{richter99}. 
Older reddening estimates range from $E(\hbox{\it B--V\/})$ = 0.32 to 0.42 
\citep{hesser76, harris79, crocker88}. 
\citet{monaco04} find that differential reddening in M22 is 
$\Delta E(\hbox{\it B--V\/})$$\sim$0.06, and our values of $E(\hbox{\it V--I\/})$ 
and $E(\hbox{\it B--V\/})$ show a similar range.
\begin{table}
\begin{scriptsize}
\centering
\caption{Reddening Determinations from RR Lyrae variables in M22}
\label{evmi}
\begin{tabular}{p{0.35in}p{0.5in}p{0.5in}p{0.5in}p{0.5in}} \\ \hline
Name & ($\hbox{\it B--V\/})_{min}$ & $E(\hbox{\it B--V\/})^b$ & ($\hbox{\it V--I\/})_{min}$ & $E(\hbox{\it V--I\/})$\\ 
\hline
V1  & 0.80$\pm$0.01 & 0.41 & 1.08$\pm$0.01 & 0.50 \\
V2  & 0.77$\pm$0.01 & 0.38 & 1.02$\pm$0.02 & 0.44 \\
V4  & 0.79$\pm$0.01 & 0.38 & 1.07$\pm$0.01 & 0.49 \\
V6  & 0.78$\pm$0.01 & 0.39 & 1.01$\pm$0.02 & 0.43 \\
V7 & 0.76$\pm$0.01 & 0.37 & 1.05$\pm$0.01 & 0.47 \\
V10 & 0.76$\pm$0.01 & 0.37 & 1.05$\pm$0.01 & 0.47 \\
V13 &  0.73$\pm$0.01 & 0.33 & 0.98$\pm$0.01 & 0.40 \\
V20 & 0.79$\pm$0.01 & 0.37 & 1.07$\pm$0.02 & 0.49  \\
V23$^a$ & 0.85$\pm$0.04 & 0.38 & 1.10$\pm$0.02 & 0.52 \\
KT-55 & 0.80$\pm$0.01 & 0.40 & 1.07$\pm$0.01 & 0.49 \\
\hline
\end{tabular}
\\
\ \ \ \ $^a$ Based on a light curve using data only from runs 9, 14, and 16, which had minimal scatter.\\
$^b$ from Equation~\ref{sturchwalker} \\ 
\end{scriptsize}
\end{table}

\citet{walker98} showed that the blue edge of the instability
strip in GCs appears to have a constant \hbox{\it B--V\/} over a wide range of metallicity,
with $(\hbox{\it B--V\/})_{0,FBE}$ = 0.18 $\pm$ 0.01.  The blue edge of our sample is 
estimated by averaging the \hbox{\it B--V\/} color of the five bluest RR Lyrae
stars and the five reddest constant stars.  
(Because measurement uncertainties guarantee observational tails beyond true limits, 
scissoring these two samples largely cancels out the systematic error that would arise
if we had used the most 
extreme \hbox{\it B--V\/} RR Lyrae to define an envelope.)
Hence the blue edge of the IS is \hbox{\it B--V\/} = 0.53 mag, suggesting
$E(\hbox{\it B--V\/})$ = 0.35 $\pm$ 0.01.  
This color excess is in excellent agreement with that found
from the RR Lyrae minimum light colors.

\clearpage

\section{The Color-Magnitude Diagram}
\begin{figure}[htb]  
\includegraphics[width=9cm]{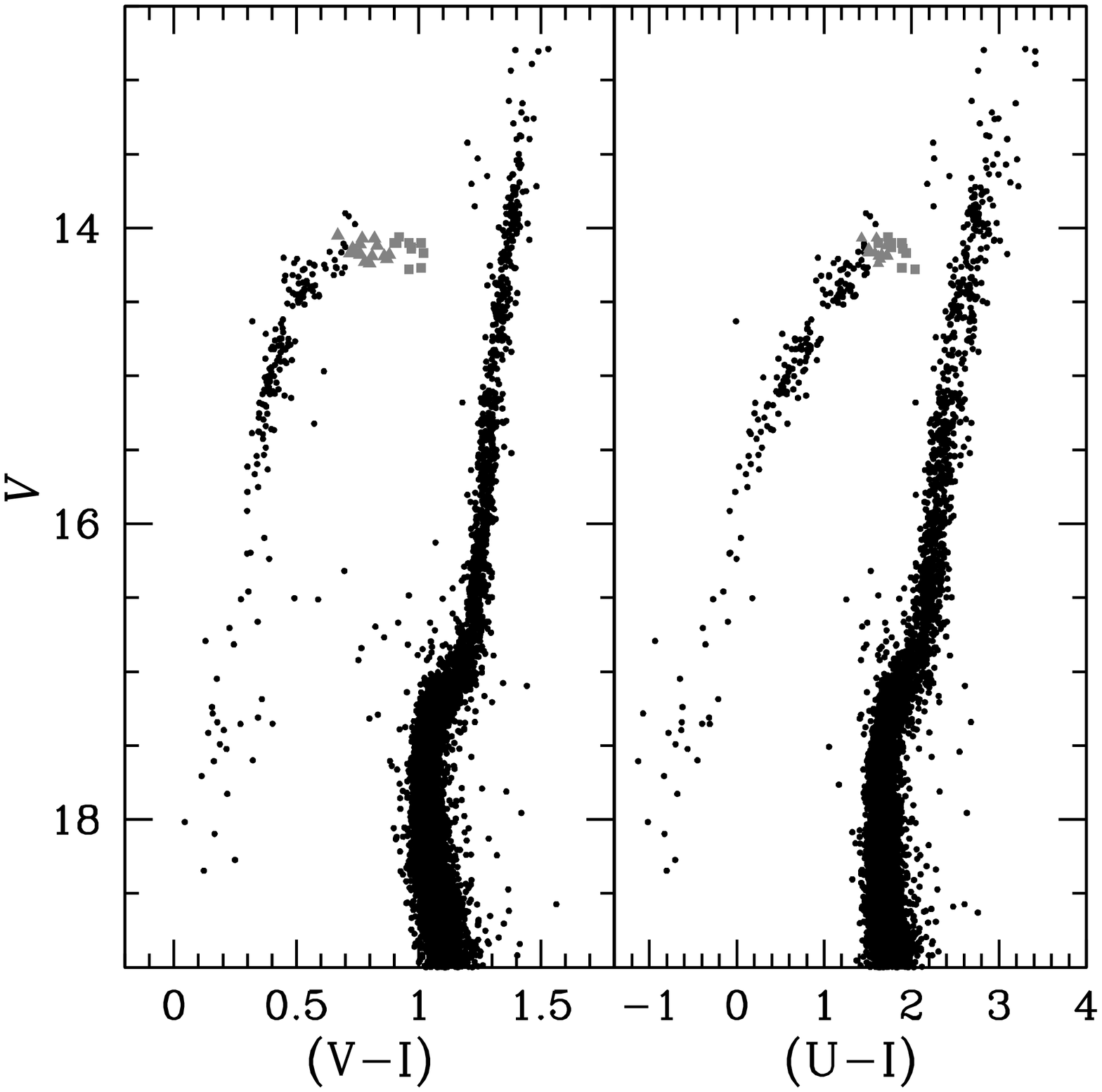}
\caption{The M22 color-magnitude diagram, showing the location of the RR Lyrae
variable stars in gray.  
\label{rrcmd}}
\end{figure}

The color-magnitude diagram for M22 is shown in Figure~\ref{rrcmd}, in $V$ 
versus \hbox{\it V--I\/} and \hbox{\it U--I\/}, where the CMD is cleaned using the
proper motions provided by \citet{zloczewski13}.  About 13\% of GCs 
in the Milky Way (15 out of 114) have strong extended HBs (E-BHBs), 
a feature which may be due to the presence of helium-enhanced second-generation 
subpopulations \citep[e.g.,][]{norris04, piotto05, lee07}.  It has also been shown that 
E-BHB GCs are in general more massive than normal GCs and have different 
kinematics \citep[e.g.,][]{lee07}.  This suggests that GCs with and without E-BHBs have 
different origins.  M22 is
known to have a strong E-BHB, and with our proper motion cleaned HB and accurate, calibrated 
photometry, we easily see that the E-BHB reaches 1.5 magnitudes fainter than the $V$-magnitude
main-sequence turnoff at its high temperature end.  

\subsection{HB morphology parameters}
Most key HB morphology parameters compare the number ratios of stars in different parts of the HB
\citep[e.g.,][]{mironov72, buonanno93, lee94}, with one of 
the most frequently used HB morphology parameters defined as
($B$-$R$)/($B$ + $V$ + $R$) \citep[see][]{lee94}.
This requires the knowledge of the number of blue ($B$) stars (bluer than the RR Lyrae gap), 
variable ($V$) and red ($R$) HB stars (number of HB stars redder than the gap) in a cluster.
For M22, $R$ is zero and $V$ is 26, as determined in \S4.  An estimate of the number of BHB, however, is 
complicated by the many field stars near M22.  The proper motion cleaned
HB ensures that the stars are cluster members, but again due to the heavy crowding of the field,
the proper motion completeness is about 50 per cent \citep{zloczewski13}.  

\begin{figure}[htb]  
\includegraphics[width=9cm]{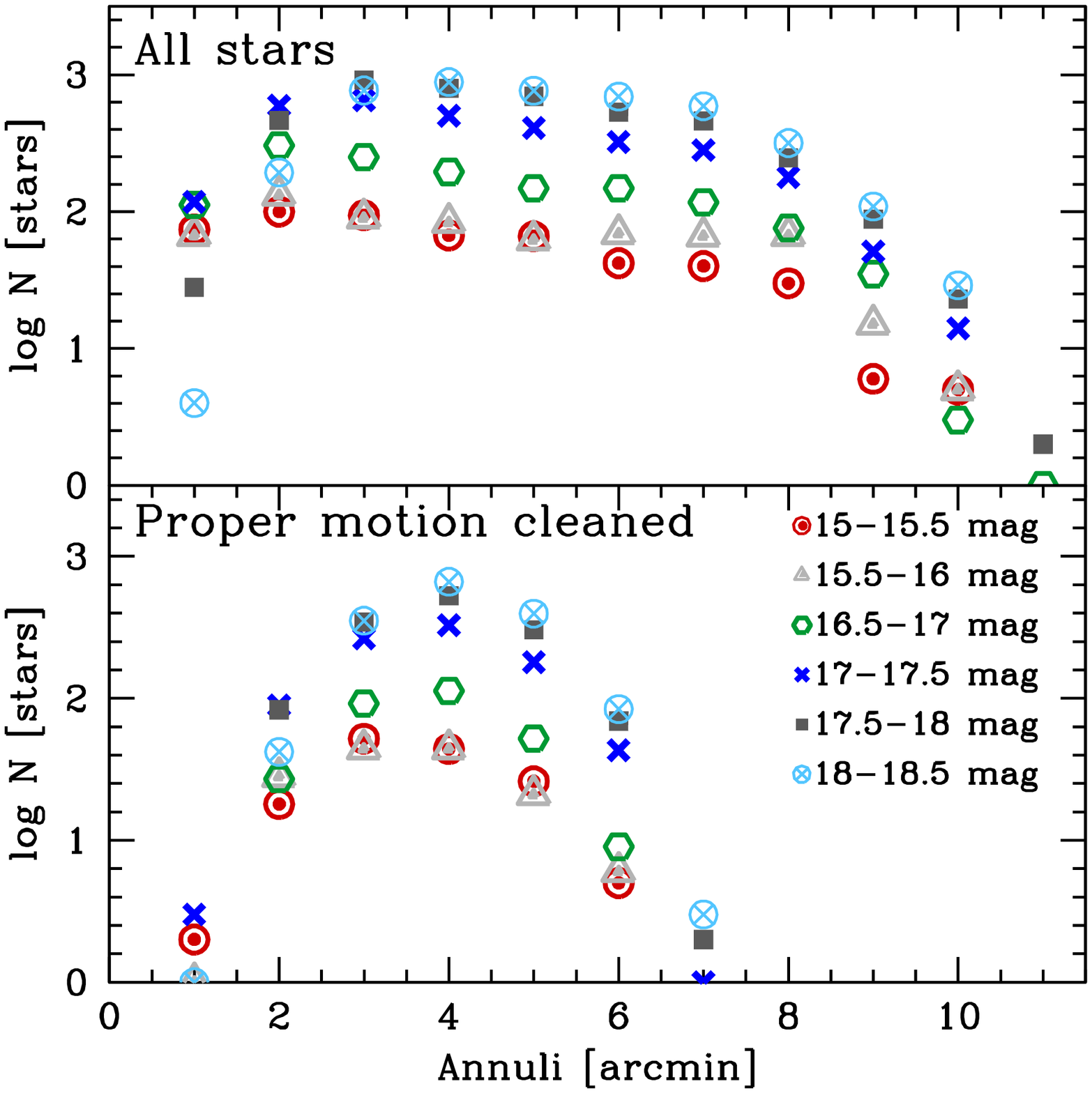}
\caption{{\it Bottom:} The number of stars per radial zone for the proper-motion selected 
M22 population. The different symbols represent different $V$ magnitudes bins, which are 
specified in the top right-hand corner.
{\it Top:}   The number of stars per radial zone for our complete sample of stars.
\label{StarCounts}}
\end{figure}

Figure~\ref{StarCounts} shows the observed star counts as a function of radial distance, and shows
the logarithm of star counts plotted against apparent magnitude.  If we can assume that the 
completeness is the same for all magnitude bins beyond some radius, say for
instance 4 arcminutes, this can be interpreted as a luminosity function for that radial zone, at least.
The radial zone that will be the most incomplete
is the innermost radial zone, where the crowding is most severe, and in the case of
the \citet{zloczewski13} proper motion 
cleaned catalog, also the outermost radial zone, where the fraction of the annulus contained
within the rectangular field of view diminishes. 
The star counts of the outermost two zones can be compared (e.g., at 6$^\prime$ and at 7$^\prime$), 
and a similar magnitude distribution would indicate they are comparably complete.  Where the magnitude 
distribution begins to diverge is where a radial zone is beginning to be incomplete.  In this way, the 
onset of where incompleteness sets in as a function of radial distance from the cluster center is 
mapped \citep[see also][]{walker11}.  

For the magnitude range of the HB, from 
$V$$\sim$15 to $V$$\sim$18.5 mag, there is no divergence in the apparent magnitude distribution from 
$r$$\sim$4-6$^\prime$ in the \citet{zloczewski13} sample.  
The number of HB stars from the \citet{zloczewski13} proper-motion selected stars between 
$r$=4$^\prime$ and $r$= 6$^\prime$ is thirty-two, and the number of proper-motion selected 
RR Lyrae stars in this same range is zero.  Therefore, 
$(B\thinspace{:}V\thinspace{:}R)$ = $\rm (32\pm6\thinspace{:}0\pm1\thinspace{:}0\pm1)$,
or ($B$-$R$)/($B$+$V$+$R$)=+1.0$\pm$0.1.

This same procedure is repeated with the sample of stars from our M22 photometric catalog.
Figure~\ref{StarCounts} shows that there is little or no sign of incompleteness at distances between
$r$=4$^\prime$ and $r$=8$^\prime$.  In this radial zone there are 147 blue HB stars, 
5 RR Lyrae stars and 0 red HB stars, giving a 
$(B\thinspace{:}V\thinspace{:}R)$ = $\rm (64\pm12\thinspace{:}2\pm2\thinspace{:}0\pm1)$
or ($B$-$R$)/($B$+$V$+$R$)=+0.97$\pm$0.1 in agreement with what was obtained from 
the proper motion cleaned sample.  

\begin{figure}[htb]  
\includegraphics[width=9cm]{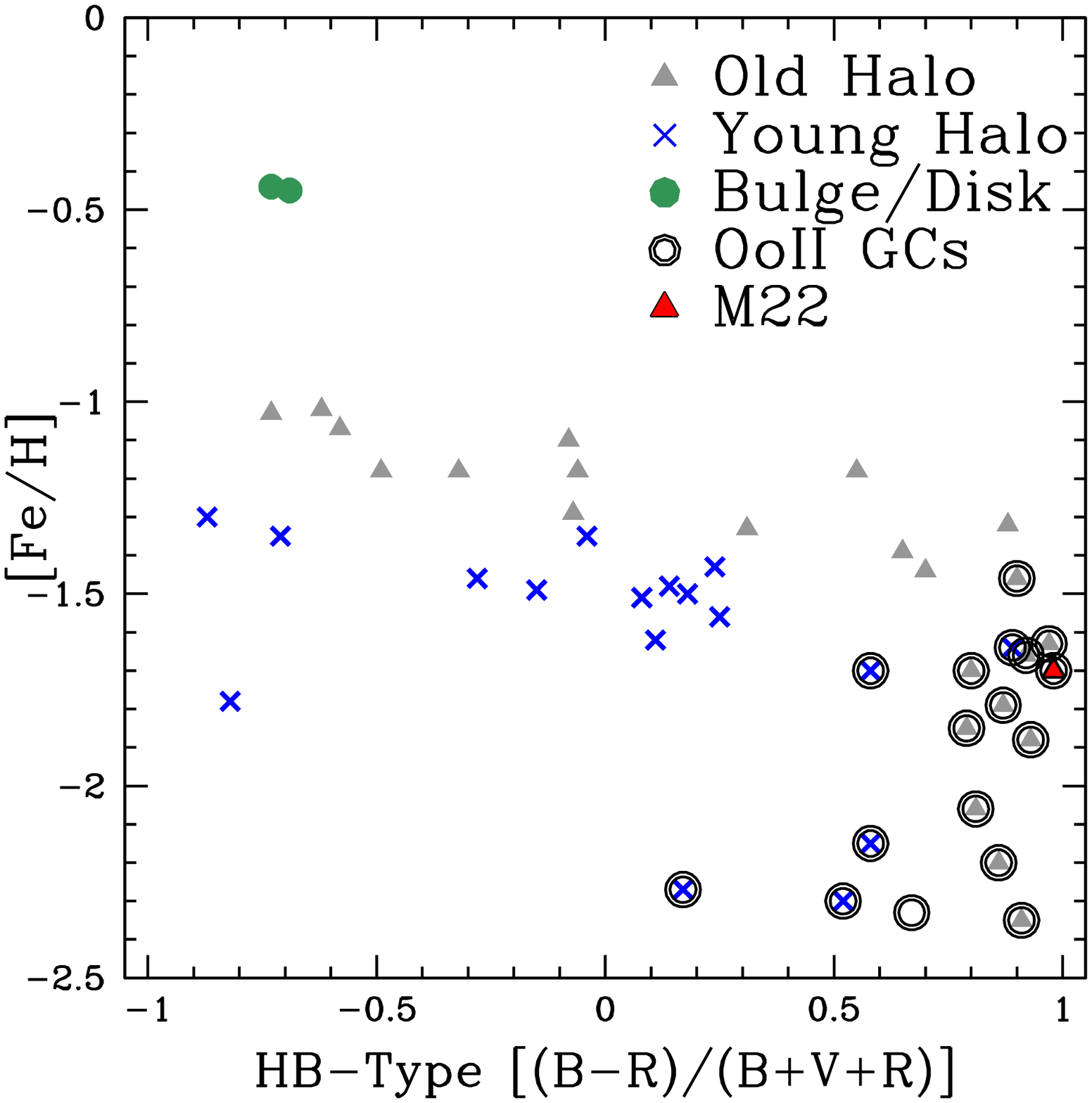}
\caption{Position of the Galactic globular clusters with a defined Oosterhoff type
in the metallicity--``HB type'' plane.  The bulge/disk, old halo and young halo clusters,
as defined by \citet{mackey05}, are labeled.
\label{HBtype}}
\end{figure}

A comparison of the HB-type of M22 with respect to other GCs that harbor a population of 
RR Lyrae variables is shown in Figure~\ref{HBtype}, where the HB-type comes from the 
compilation of \citet{catelan09}.  As expected, M22 has a $\rm [Fe/H]$, HB-type and Oosterhoff 
classification consistent with an old halo GC.


\subsection{Gaps on the Blue HB}
\begin{figure}[htb]  
\includegraphics[width=9cm]{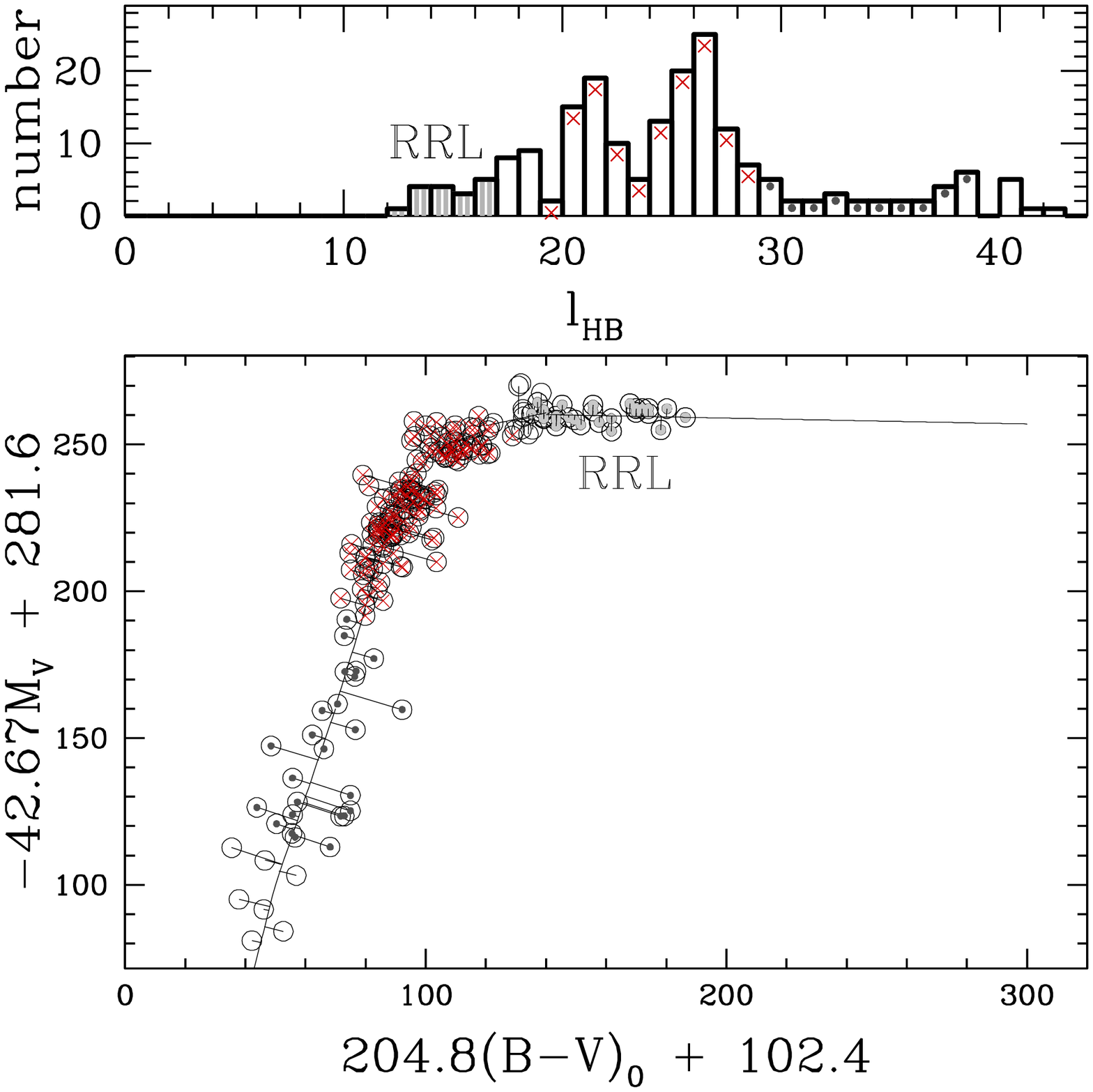}
\caption{{\it Bottom:}  The projection of the proper motion cleaned HB stars of M22 
onto the HB ridge-line, with the adopted ridge-line and projection vector of each star shown. 
Using this scale, one unit in the abscissa has the same length as one unit in the 
ordinate \citep[see][for details]{piotto99}. 
At 204.8$(\hbox{\it B--V\/})_0$ + 102.4 = 300 is the adopted zero point for $l_{HB}$ and
corresponds to \hbox{\it B--V\/} = 0.965.
Different symbols indicate the RR Lyrae variables, as well as the stars with $l_{HB}$ = 0, 10, 
20, 30, 40 along curve.
{\it Top:} The observed distribution of stars along the HB.
\label{lhb}}
\end{figure}

Rood \& Crocker (1985, 1989) suggest that a good method for analyzing the distribution 
of stars along the HB is to define a coordinate, $l_{HB}$, which is linear along the HB ridge-line.
Briefly, the length of the HB ridge-line is divided into equal bins, and the number of stars populating
each bin (perpendicular to the ridge-line) is determined (see bottom of Figure~\ref{lhb}).  
Although $l_{HB}$ has been used frequently \citep[e.g.,][]{ferraro92, dixon96, catelan98},
it has been given different definitions (i.e., different bin sizes along the HB ridgeline), 
making it difficult to accurately reproduce and compare from cluster to cluster.  In response
to such confusion, \citet{piotto99} outline a recipe to obtain a ``standard" $l_{HB}$, and 
their procedure is adopted here.  

The $l_{HB}$ distribution is shown in Figure~\ref{lhb}, using the proper motion cleaned 
HB stars as well as the RR Lyrae stars in the cluster.  The $l_{HB}$ distribution is asymmetric;
that the stellar distribution along the HB is not symmetric is a feature characteristic of 
blue HB clusters.  The main feature of the $l_{HB}$ is a 
double peaked distribution, one peak at $l_{HB}$=26 and another at $l_{HB}$=21.  
Between the two peaks is a pronounced gap, at $l_{HB}$=23, clearly seen at $V$= 14.6,
and first noted by \citet{cho98}.  To our knowledge, this is the first time that a gap clearly 
shows up in the middle of the so-called hot HB stars. 

Unfortunately, it is difficult to tie physical parameters 
(such as $\rm T_{eff}$ or mass) to any gaps or over-densities along the HB from 
the $l_{HB}$ parameter.  This is because the HB ridge line does not follow the ZAHB, 
whereas theoretical models do.  We have therefore used the ZAHB shown in 
Figure~\ref{cmdhb} to estimate the temperature of where this gap occurs.  In particular,
the ZAHB with a normal CNO and $\rm [Fe/H]$=$-$1.84 dex is used, as that
has spectroscopically been shown to be appropriate for the M22 HB \citep{marino13}.  
From (\hbox{\it V--I\/},$V$)=(0.44,14.53) to (\hbox{\it V--I\/},$V$)=(0.41,14.71), which is 
the approximate range of the ``gap" in a  $V$ versus \hbox{\it V--I\/} (see Figure~\ref{rrcmd}),
the $T_{eff}$ runs from 9,500 K to 10,500 K.  Therefore, the HB gap in M22 
is on the cool side of the \citet{grundahl98} ``jump", which is seen in the UV CMDs of E-BHB GCs 
at $T_{eff}^{jump}$=11,500 $\pm$ 500 K \citep[e.g.,][]{grundahl99} .

A peak at $l_{HB}$=26 is also seen in NGC 6273 \citep{piotto99}, suggesting that there may
be some preference for stars to clump here, although a sample of two clusters is not sufficient to draw 
any conclusions at this point.  


%
\subsection{Radial Distribution}
The currently available models for the formation of multiple populations in a GC
suggest that the second generation stars should form in the inner regions of the cluster, 
shortly after the ejecta from the more massive first generation stars accumulated there and 
subsequently mixed with pristine, unpolluted matter \citep[e.g.,][]{dercole08, decressin10}.  
This is actually observed in the case of the not-yet dynamically
relaxed GCs $\omega$$\,$Cen  and M13 \citep{sollima07, bellini09, lardo11, johnson12}. 
Despite these cases, it is believed that the majority of the MW GCs will have the presence of 
radial stellar population gradients erased as a consequence of the long-term dynamical 
evolution occurring during  the last $\sim$ 12 Gyr of their life \citep[e.g., as it seems is the 
case for the GC NGC$\,$1851;][]{milone09}.

\begin{figure}[htb]  
\includegraphics[width=9cm]{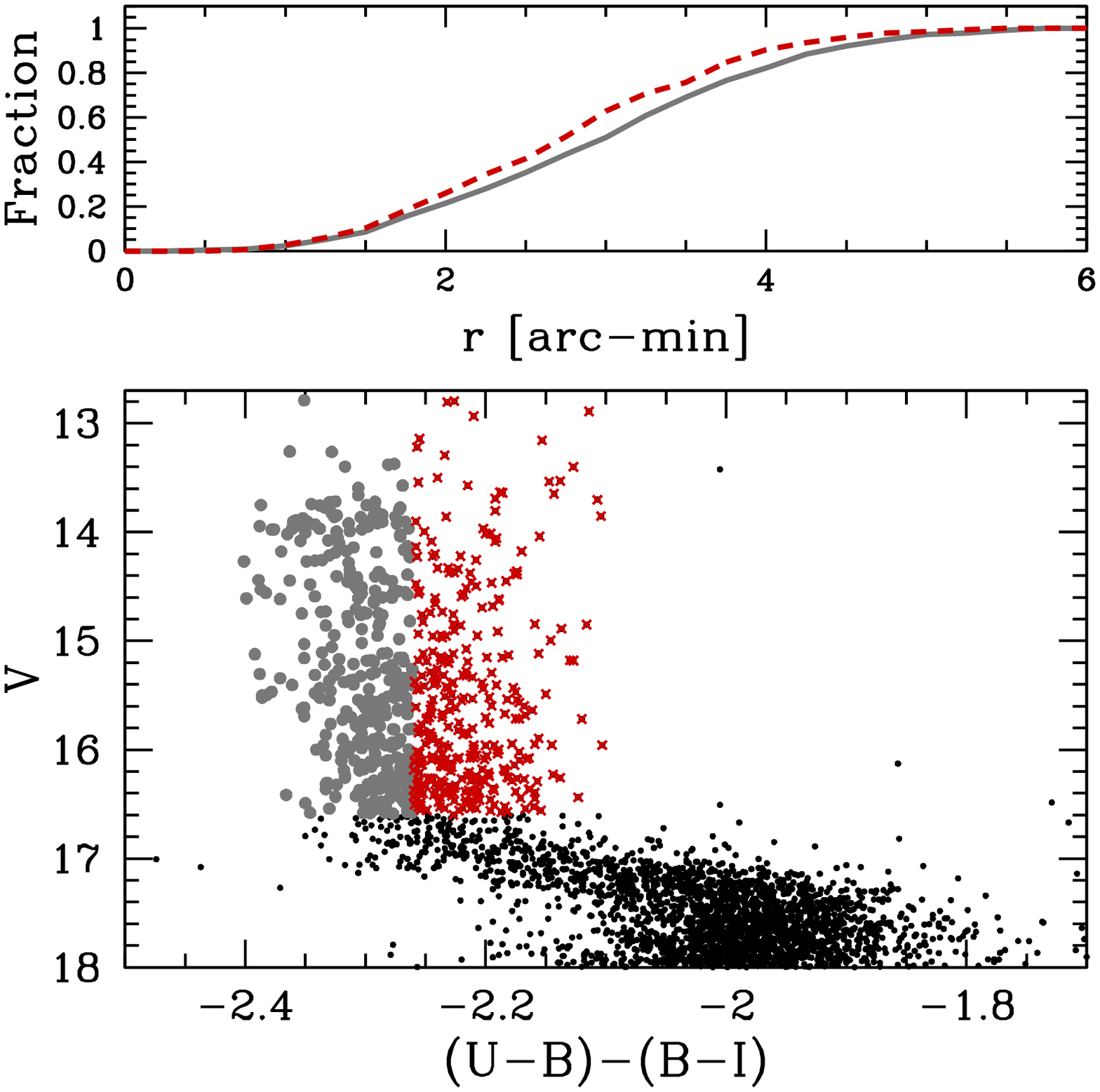}
\caption{
The RGB of M22 in the $V$ vs. $c_{U,B,I}$ diagram, where 
$c_{U,B,I}$ = (\hbox{\it U--B\/})-(\hbox{\it B--I\/}) is sensitive to stellar helium and 
light element content.  The grey and red points are arbitrarily divided at 
$c_{U,B,I} = 2.26$ to separate blue and red RGB stars, respectively.  (Top): The 
cumulative radial distributions of red RGB (dashed curve) and blue RGB stars (solid curve).
%
\label{rgb_pm2}}
\end{figure}

That M22 has a double RGB with different Ca abundances has been shown from 
by \citet{lee09} from the $hk$ index of the $Ca$-$by$ photometry.  \citet{joo13} also 
use Str\"{o}mgren photometry to show that the RGB consists of two stellar populations.  
In Figure~\ref{rgb_pm2} we use the $V$ vs $c_{U,B,I}$ diagram, where 
$c_{U,B,I}$ = (\hbox{\it U--B\/})-(\hbox{\it B--I\/}), to separate the blue and red RGB stars.  
Here, the red RGB stars are arbitrarily defined as those with $c_{U,B,I}$ $<$ $-$2.26
and the blue RGB stars have $c_{U,B,I}$ $>$ $-$2.26.
The $c_{U,B,I}$ index was introduced by \citet{monelli13} and maximizes the separation 
among stars with different helium and light-elements content.  This is because 
the \hbox{\it U--B\/} color is sensitive to light-element variations
\citep{marino08, sbordone11} and the \hbox{\it B--I\/} color is very efficient in disentangling stellar
populations with different helium abundance \citep{piotto07, dicriscienzo11}.
The radial distributions of the blue and red RGB stars are shown in the top
panel of Figure~\ref{rgb_pm2}.  
and suggests that the red population is slightly 
more centrally concentrated than the blue RGB population.  This is also confirmed from 
the results of a two-sided KS test, which indicates that the red
and blue RGB stars trace the same radial distribution with KS-prob = 0.10.  Such
a KS-prob is just marginally greater than the default threshold KS-prob = 0.05 
below which one  rejects the null hypothesis.

\begin{figure}[htb]  
\includegraphics[width=9cm]{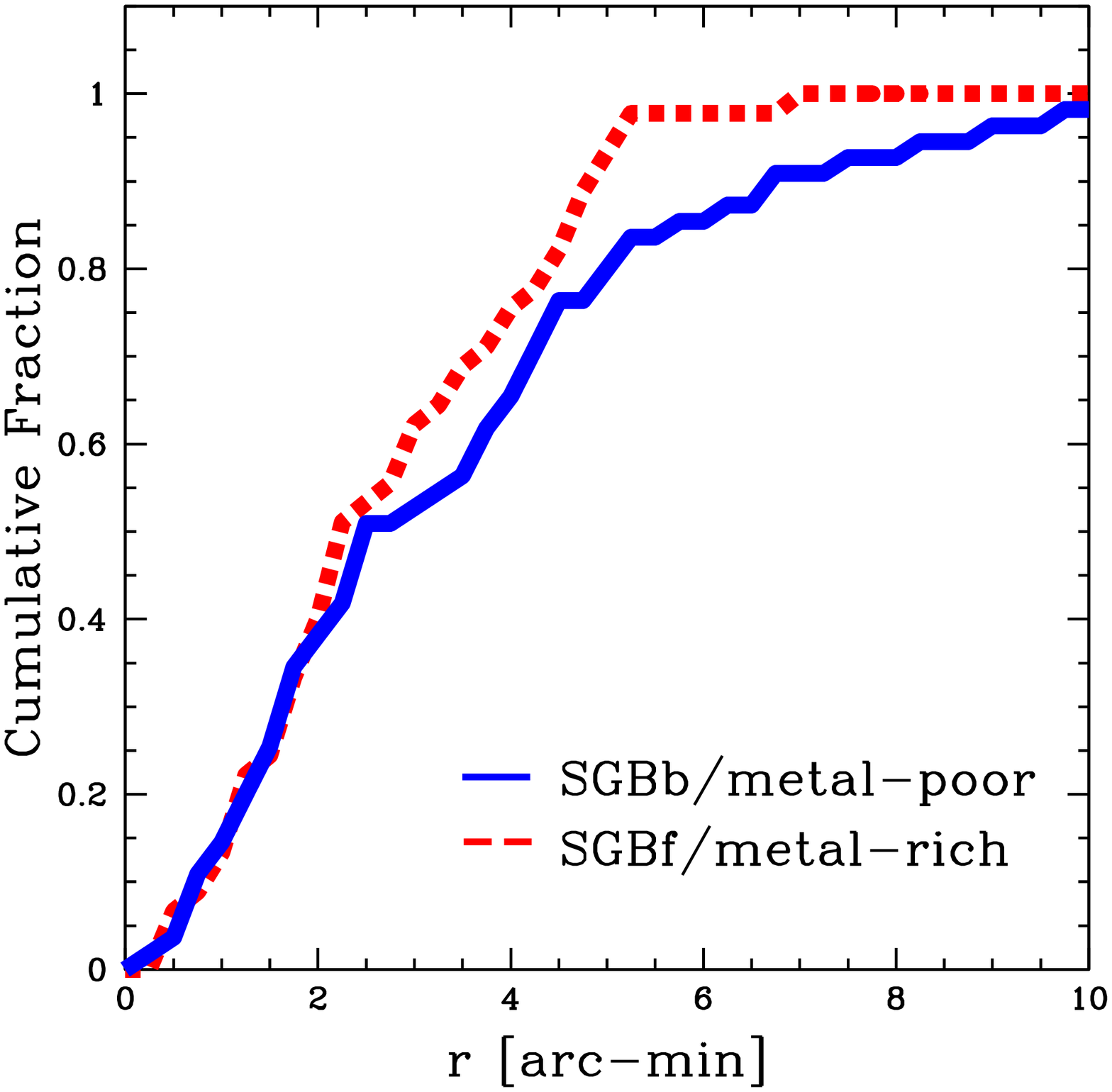}
\caption{Normalized cumulative distribution of the \citet{marino12} sample of SGBb and SGBf stars
as a function of distance from the cluster center. 
\label{sgb_rad}}
\end{figure}

Similarly, \citet{marino12} present an abundance analysis of 101 SGB stars in M22, 
finding that the faint SGB is populated by more metal-rich stars than the bright SGB.  
\citet{marino12} classify their SGB stars as $s$-poor and $s$-rich based on the 
spectroscopic C, Sr and Ba abundances, and the cumulative radial distribution of 
their sample of SGB stars (their Table~3) is shown in Figure~\ref{sgb_rad}.  
Although the sample is small, the $s$-poor and $s$-rich stars appear to have different 
radial distributions beyond $\sim$3 arc minutes.
The result of a two-sided KS test indicates that the faint SGB and bright SGB stars trace 
the same radial distribution with KS-prob = 0.19 beyond a radius of 3 arc minutes.  

The faint SGB evolves in a 
redder RGB sequence whereas the brighter SGB evolves in a bluer branch.
Both the RGB (red/blue) and SGB (bright/faint) populations show hints of radial
trends, but ultimately, cleaner and larger samples are needed to make firm conclusions
of the long-term dynamical evolution of M22.

In general, GCs with strong E-BHBs, as seen in the HB of M22, are modeled by 
helium-enhanced second-generation subpopulations \citep[e.g.,][]{norris04, piotto05, lee07}.
As the second generation of stars
in M22 can be distinguished radially, as shown clearly with the M22 SGB and RGB stars, 
it is of interest to use radial plots to search for trends in the stars along the HB.  
%

\begin{figure}[htb]  
\includegraphics[width=9cm]{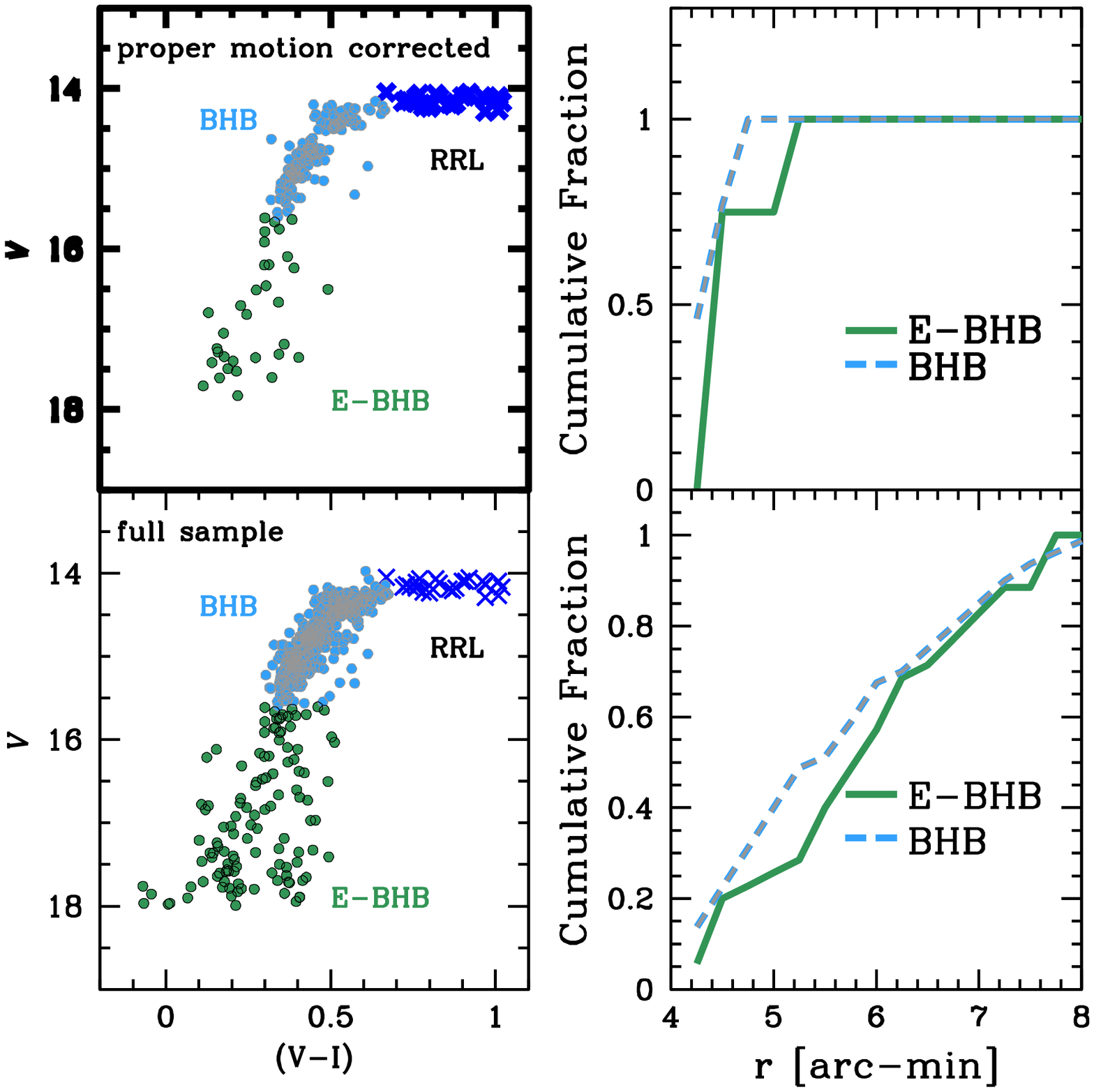}
\caption{Normalized cumulative distribution of the E-BHB stars and the BHB stars
as a function of distance from the cluster center. 
\label{hb2_rad}}
\end{figure}

Figure~\ref{hb2_rad} shows the radial distribution of the E-BHB stars, using our photometry
and also using the proper-motion selected stars from \citet{zloczewski13}.  We limit our 
analysis to the stars at distances between $r$=4$^\prime$ and $r$=6$^\prime$ and 
to $V$=18.0 mag, as we believe that this is the range where these samples show no 
magnitude bias (see Figure~\ref{StarCounts} and discussion).  
Both our sample and the \citet{zloczewski13} sample show that the E-BHB 
stars can not be distinguished from the rest of the HB stars in a radial plot.
Unfortunately, the number of HB stars in the outer parts of M22 is significantly smaller than 
the number of stars closer to the central part of the cluster.  For example, there are only 
36 E-BHB stars in our sample at distances larger than $r$=4$^\prime$ from the cluster center.  
Therefore the significance of this finding might be somewhat biased.  If confirmed, 
however, this would indicate that the most extreme BHB stars in M22 are formed 
alongside the reddest part of the HB.  Because \citet{marino13} have shown that 
the reddest HB stars belong to the first generation of stars, this would
then also suggest that the E-BHB stars in M22 belong to the first generation and are
not He enhanced.

Our result of no strong radial separation of the HB components in M22 is in agreement 
with that found in the globular cluster NGC$\,$2808 \citep{iannicola09}.  They divide the 
NGC$\,$2808 HB stars into three radial bins, and find that the relative fractions of cool, 
hot, and extreme HB stars do not change when moving from the center to the outskirts of 
the cluster.  
It was also found from high resolution spectra of hot HB stars in 
$\omega$ Cen, that a significant fraction of E-BHB stars are helium-poor \citep{moehler11}.
They find that among the hottest stars in the E-BHB (in the temperature range 
30 000 K to 50 000 K), $\sim$30\% are actually helium-poor.

Our result is also in agreement with some of the M22 population models presented by
\citet{joo13} (see their Figure~13).  For example, \citet{joo13} show that if
there is a difference in $\rm [CNO/Fe]$ between the two SGB M22 subpopulations, 
as suggested for NGC$\,$1851 \citep{cassisi08, salaris08}, both the E-BHB and the
reddest HB will have a roughly equal mix of first and second generation stars. 
Additional theoretical models by \citet{moehler11} have shown that the presence of 
E-BHB stars can be explained 
as having formed by an independent evolutionary channel: hot helium flashers
and a significant fraction of E-BHB stars may come from the above channel.
Therefore, here we present additional observational results in contradiction to the 
generally accepted view on the formation of an E-BHB
\citep[e.g.,][]{norris04, piotto05, lee07, dantona08}.  Especially in this regard, radial studies 
involving wide-field photometry on a large sample of E-BHB clusters would be a 
worthwhile endeavor.

\section{Conclusions}
The first calibrated {\it UBVI\/} photometry is presented for 26 RR Lyrae variables in the chemically 
heterogeneous, split-SGB globular cluster NGC$\,$6656 (M22).  Four of these variables are newly
discovered, which increases the specific RR Lyrae fraction of M22 to $S_{RR}$=10.4 where 
$S_{RR} = N_{RR} \times 10^{0.4(7.5+M_V)}$, and $M_V$=$-$8.50 \citep[][2010 update]{harris96} 
is the cluster's integrated absolute magnitude in $V$.  The pulsational parameters of the variables 
suggest that 10 RR Lyrae variables are fundamental-mode pulsators with mean periods of 
$\rm \langle P_0\rangle$ = 0.66$\pm$0.02 days and 16 are first-overtone variables with mean 
periods of $\rm \langle P_1\rangle$= 0.33$\pm$0.01 days.  The number ratio of the RR$1$-type 
variables to the total number of the RRL-type variables is $N_1/N_{RR}$ = 0.61. 
Therefore M22 can be classified as an OoII cluster, which is expected considering the 
metallicity and horizontal branch morphology of the cluster.

The mean magnitude for the RRLs is $\rm m_{V,RR}$ = 14.15 $\pm$ 0.02 mag,
and the reddening from their minimum light colors was determined to be 
$E(\hbox{\it B--V\/})$=0.36$\pm$0.02 mag, in agreement with previous 
studies \citep[e.g.,][]{monaco04, richter99}.  Using both the recent recalibration of 
the RR Lyrae luminosity scale by \citet{catelancortes08} and the quadratic relation between 
RR Lyrae absolute magnitude and metallicity from Bono, Caputo, \& di Criscienzo (2007),
the RR Lyrae variables have absolute magnitudes of $M_V$=0.57$\pm$0.13.  
This leads to an RR Lyrae distance of 
$(m-M)_{V,RRL}$=13.58$\pm$0.13 and adopting $E(\hbox{\it B--V\/})$=0.36 mag, 
$(m-M)_{0,RRL}$=12.46$\pm$0.13 mag. 
We note, however, that a 70\% brighter $M_V$ is found when using the \citet{benedict11}
$M_V$--$\rm [Fe/H]$ relation, which would change the RR Lyrae distance modulus by $\sim$0.2 mag.
A somewhat brighter $M_V$ is in accordance to the distance determination of $(m-M)_{V}$=13.65~mag 
found by using theoretical Zero Age Horizontal Branch loci and accounting for the most recent 
spectroscopic analysis of the chemical composition of the cluster stars.

From the comparison of theoretical ZAHB sequences with observations, we find
that most RRLs have a mass of M=0.64--0.68$M_\sun$ if there is no non-canonical enhancement of 
CNO elements or helium.  If such non-canonical enhancement {\it is\/} present, the mass range
could be more like 0.60--0.61$M_\sun$.  We also find that, theoretically,
if two populations of RR Lyrae variables existed in M22 with abundance ratios similar to
the two populations of SGB stars \citep[e.g.,][]{marino12}, it would be very difficult to distinguish
between the more metal-rich and more metal-poor variables from their luminosities alone.
This is because the $s$-rich stars have higher C and N contents, so
their luminosities would be comparable to the more metal-poor variables.


After correcting for completeness on the horizontal branch, 
 ($B$-$R$)/($B$+$V$+$R$)=+0.97$\pm$0.1 is obtained.  This is one of the largest HB values
for a GC with a substantial population of RR Lyrae stars, and combined with its $\rm [Fe/H]$
and Oosterhoff type, is consistent with the classification of an old halo GC.

Different spatial distributions can be separated in the \citet{marino12} sample of SGB stars as well 
as in the RGB, with the more metal-rich SGBf and the redder RGB being marginally more centrally 
concentrated than the SGBb and bluer RGB.  Therefore a search of different radial distributions 
in the HB stars has been carried out.  Although the E-BHB stars are usually explained as 
being a helium-enhanced second-generation subpopulation \citep[e.g.,][]{norris04, piotto05, lee07},
we are not able to radially distinguish the E-BHB from the rest of the stars on the horizontal branch.
This may mean that the E-BHB stars are not exclusively a second-generation subpopulation, 
as also seen with the E-BHB of NGC$\,$2808 \citep{iannicola09} and the E-BHB of
$\omega$ Cen \citep{moehler11}.  However, the small numbers of E-BHB stars at large cluster radii 
limit the significance of this interpretation.

Combining our photometry with the proper-motion cleaned sample by \citet{zloczewski13}, 
we find at least one ``gap" and two over-densities on the blue HB.  Although the gap is 
located in an unusual part of the HB, at $V$=14.6 mag and in the middle of the so-called 
hot HB stars, the over-density at $l_{HB}$=26 is also seen in NGC 6273 \citep{piotto99}, 
suggesting that there may be some preference for stars to clump here.  


\acknowledgments
This research uses services or data provided by the NOAO Science Archive. NOAO is operated 
by the Association of Universities for Research in Astronomy (AURA), Inc. under a cooperative 
agreement with the National Science Foundation.  This research is also based on 
observations obtained at the Southern Astrophysical Research (SOAR) telescope, 
which is a joint project of the Minist\'{e}rio da Ci\^{e}ncia, Tecnologia, e Inova\c{c}\~{a}o (MCTI) 
da Rep\'{u}blica Federativa do Brasil, the U.S. National Optical Astronomy Observatory (NOAO), 
the University of North Carolina at Chapel Hill (UNC), and Michigan State University (MSU).
G.B. thanks ESO for support as a science visitor. This work was partially supported by 
PRIN-INAF 2011, (P.I.: M. Marconi) and by PRIN/MIUR (2010LY5N2T), (P.I.: F. Matteucci).
J.-W.L. acknowledges financial support from the Basic Science
Research Program (grant No. 2010-0024954) and the Center
for Galaxy Evolution Research through the National Research
Foundation of Korea.
Support for M.C. is provided by the Chilean Ministry for the Economy, Development, and 
Tourism's Programa Iniciativa Cient\'{i}fica Milenio through grant P07-021-F, awarded to The 
Milky Way Millennium Nucleus; by the BASAL Center for Astrophysics and Associated 
Technologies (PFB-06); by Proyecto Fondecyt Regular \#1110326; and by Proyecto Anillo ACT-86.

\clearpage

\clearpage

\end{document}